\documentclass[12]{article}

\usepackage{amsmath,amssymb,amsthm}
\usepackage{hyperref}
\usepackage{natbib}
\usepackage{mathtools}
\usepackage[title]{appendix}
\usepackage{fullpage}
\usepackage{setspace,rotating}
\usepackage{threeparttable,booktabs}
\usepackage[utf8]{inputenc}

\newtheorem{theorem}{Theorem}

\newtheorem{assumption}{Assumption}

\theoremstyle{remark}

\newcommand{\E}{\mathrm{E}}
\newcommand{\Var}{\mathrm{Var}}
\newcommand{\Cov}{\mathrm{Cov}}

\hypersetup{
	colorlinks   = true, 
	urlcolor     = blue, 
	linkcolor    = blue, 
	citecolor   = blue 
}

\begin{document}
	
	\title{The Yule-Frisch-Waugh-Lovell Theorem for Linear Instrumental Variables Estimation}
	
	\author{Deepankar Basu\thanks{Department of Economics, University of Massachusetts Amherst. Email: \texttt{dbasu@econs.umass.edu}. Without implicating them in any way in the contents of this paper, I would like to thank Michael Ash, Kit Baum, Sergio Correia, William Greene and James MacKinnon for helpful comments on an earlier version of this paper.} }
	
	\date{\today}
	
	\maketitle
	
	\begin{abstract}
		In this paper, I discuss three aspects of the Frisch-Waugh-Lovell theorem. First, I show that the theorem holds for linear instrumental variables estimation of a multiple regression model that is either exactly or overidentified. I show that with linear instrumental variables estimation: (a) coefficients on endogenous variables are identical in full and partial (or residualized) regressions; (b) residual vectors are identical for full and partial regressions; and (c) estimated covariance matrices of the coefficient vectors from full and partial regressions are equal (up to a degree of freedom correction) if the estimator of the error vector is a function only of the residual vectors and does not use any information about the covariate matrix other than its dimensions. While estimation of the full model uses the full set of instrumental variables, estimation of the partial model uses the residualized version of the same set of instrumental variables, with residualization carried out with respect to the set of exogenous variables. Second, I show that: (a) the theorem applies in large samples to the K-class of estimators, including the limited information maximum likelihood (LIML) estimator, and (b) the theorem does not apply in general to linear GMM estimators, but it does apply to the two step optimal linear GMM estimator. Third, I trace the historical and analytical development of the theorem and suggest that it be renamed as the Yule-Frisch-Waugh-Lovell (YFWL) theorem to recognize the pioneering contribution of the statistician G. Udny Yule in its development. 
		
	\end{abstract}
	
	\doublespacing
	
	\section{Introduction}
	
	The Frisch-Waugh-Lovell (FWL) theorem is a remarkable result about linear regression models estimated with the method of least squares. The theorem shows that coefficients of variables in a multiple regression are exactly equal to corresponding coefficients in partial regressions that use residualized versions of the dependent and independent variables. While it has been primarily used in econometrics \citep{krishnakumar-2006,tielens-vanhove-2017}, it has found wider applications in a variety of disciplines, including statistics \citep{arendacka-puntanen-2015,cinelli-hazlett-2020,gross-moeller-2023}, electrical engineering \citep{monsurro-trifiletti-2017}, computer science \citep{ahrens-etal-2021}, and genetics \& molecular biology \citep{reeves-etal-2012}, to name just a few. It is now included in many graduate-level textbooks in econometrics, including \citet[p.~19--24]{davidson-mackinnon}, \citet[p.~73--74]{hayashi}, \citep[p.~62--69]{davidson-mackinnon-2004}, \citet[pp.~35--36]{angrist-pischke-2009}, and \citep[p.~33]{greene}. The theorem is widely used for estimating linear regression models with large number of dummy variables by partialiing them out, e.g. in panel data sets with large number of fixed effects \citep{mccaffrey-etal-2012, gaure-2013, Correia2017:HDFE}.  
	
	The first contribution of this paper is to show that the FWL theorem can be derived for generalized linear instrumental variables estimation. In particular, consider a linear regression model with a mix of endogenous and exogenous variables. Suppose a researcher has a set of valid instrumental variables, where the number of instrumental variables is weakly larger than the number of endogenous variables---thus I allow for both exactly and over-identified models. Now consider two models: a full model and a partial model. In the full model, the researcher estimates all the parameters of the regression function using instrumental variables estimation (with the full set of instrumental variables). In the partial model, the researcher first residualizes the outcome variable, the endogenous regressors and the full set of instrumental variables with the set of exogenous variables, and then estimates a linear regression of the residualized outcome variable on the residualizes set of endogenous variables using instrumental variables estimation with the residualized instrumental variables. 
	
	In such a context, I show that: (a) the coefficient vector on the endogenous variables are numerically identical for instrumental variables estimation of the full and the partial model; (b) the vector of residuals are numerically identical for instrumental variables estimation of the full and partial model; and (c) estimated covariance matrices of the coefficient vector on the endogenous variables from the full and partial models are identical (up to a degree of freedom adjustment) if the estimator of the vector of regression errors are only functions of the regression residuals (and the only information used from the regressor matrices are their dimensions). Hence, this is the FWL theorem for instrumental variables estimation. This implies, in particular, that the argument about computational efficiency that led to widespread use of the FWL in least squares estimation of linear models with large number of dummy variables \citep{mccaffrey-etal-2012, gaure-2013, Correia2017:HDFE}, can also be applied to instrumental variables estimation of such models.
	
	The second contribution of the paper is to investigate if the FWL theorem applies to two closely related estimators: the limited information maximum likelihood (LIML) estimator and the linear generalized method of moments (GMM) estimator. I show that the FWL theorem applies in large samples to all K-class estimators (the LIML estimator being an important member of this class) for which the parameter $K$ that defines the estimator, tends to $1$. I also show that the theorem does not apply to linear GMM estimators (for overidentified models) in general, but that it does apply to the two step optimal linear GMM (2SGMM) estimator. The reason for this has to do with whether the estimator in question has the same structure as the 2SLS estimator, including, most importantly, in its expression a projection matrix.\footnote{Any square matrix which is symmetric and idempotent is known as a projection matrix \citep[p.~164]{strang-2006}.} I show that it is possible to express all K-class estimators, where $K \to 1$ with sample size, in terms of an approximate projection matrix. In a similar manner, I also show that it is possible to express the two step optimal GMM estimator in terms of a suitably defined projection matrix. That is why the FWL theorem applies, approximately, to all consistent K-class estimators, including the LIML estimator, and equally well applies to the 2SGMM estimator. I give an empirical example to illustrate this point for both LIML and the optimal 2SGMM estimators. 
	
	The third contribution of the paper is to offer a historical narrative about the analytical development of the FWL theorem. Through this analytical narrative, I wish to highlight two points. The first point relates to the contribution of the statistician G. Udny Yule in the development of the FWL theorem. In the econometrics literature, the theorem is understood to have originated in a $ 1933 $ paper by Ragnar Frisch and Frederick V. Waugh in the first volume of  \textit{Econometrica} \citep{frisch-waugh-1933}, which was later generalized by Michael C. Lovell \citep{lovell-1963}. In fact, the result was proved more than two and a half decade ago by Yule in a $ 1907 $ paper \citep{udny-1907}. This seems to be well known in statistics \citep[p.~60]{agresti-2015} and should be recognized in econometrics as well. To recognize Yule's pioneering contribution to the development of this important result, I suggest that we refer to it as the Yule-Frisch-Waugh-Lovell (YFWL) theorem, rather than the currently used Frisch-Waugh-Lovell (FWL) theorem. 
	
	The second point is to trace out the analytical \textit{development} of the theorem through the decades highlighting the contribution at each stage of its development. In \citet{udny-1907}, who proved the result using basic algebra, the partial regressions are always bivariate regressions of residual vectors. In \citet{frisch-waugh-1933}, where the proof relied on some basic properties of determinants and Cramer's rule for finding solutions of linear systems of equations, the partial regressions are themselves multiple regressions but the residuals are computed with bivariate regressions. Thus, \citet{udny-1907} allows multiple variables in the conditioning set, but conceives of the partial regressions as bivariate regressions. On the other hand, \citet{frisch-waugh-1933} allows only one variable (the linear time trend) in the conditioning set, but allows the partial regressions to be multiple regressions. 
	
	Thirty years later, \citet{lovell-1963} generalizes the theorem significantly. He allows multiple variables in the conditioning set and allows the partial regressions to themselves be multiple regressions.\footnote{\citet[p.~1000]{lovell-1963} points out that \citet[p.~304]{tintner-1957} had extended the Frisch-Waugh result to the case of polynomial time trends.} In terms of methodology, \citet{lovell-1963} introduces the use of projection matrices from linear algebra, making the proof compact and elegant. Current presentations of the theorem closely follow Lovell's exposition \citep{davidson-mackinnon-2004,greene}. Finally, \citet{lovell-1963} shows that the vector of residuals from the full and partial regressions are identical, a result that neither \citet{udny-1907} nor \citet{frisch-waugh-1933} had investigated. This finding led Lovell to comment briefly on degrees of freedom adjustment necessary for statistical inference.
	
	Till the 1960s, the YFWL theorem was used in the context of linear regression models estimated with least squares. While \citet{udny-1907} and \citet{frisch-waugh-1933} had considered ordinary least squares (OLS) estimation, \citet{lovell-1963} had pointed out that the theorem could be extended easily to generalized least squares estimation. Could the same set of results be derived for instrumental variables estimation? \citet{giles-1984} provided an answer in the affirmative for an exactly identified model (where the number of instrumental variables are exactly equal to the number of endogenous variables in the model). \citet{giles-1984} showed that both the coefficient vector and the residuals would be identical from instrumental variables estimations of the full and partial models.
	
	The issue of statistical inference had not yet been investigated thoroughly, i.e. would researchers be able to conduct inference about the subset of coefficients in a multiple regression using the standard errors (or the covariance matrix) estimated from the partial regression? It is true that \citet{lovell-1963} had addressed this issue briefly by pointing out that \textit{estimated} standard errors from the multiple and partial regressions are equal up to a degree of freedom adjustment when the true error is homoskedastic. But he left it at that. It is only in \citet{ding-2021} that we find a systematic treatment of the issue. \citet{ding-2021} has demonstrated that, in the context of least squares estimation, estimates of the covariance matrices from the multiple and partial regressions are equal up to a degree of freedom adjustment in the case of homoskedastic errors and, quite surprisingly, are exactly equal for some variants of heteroskedasticity consistent covariance matrices (HCCM), heteroskedasticity and autocorrelation consistent (HAC) covariance matrices, and for some variants of clustered robust covariance matrices. 
	
	In fact, the analysis in \citet{ding-2021} highlights a general principle: if the estimate of error covariance matrix is a function only of the residuals and does not depend on the matrix of covariates beyond its dimensions, then estimated covariance matrices from multiple and partial regressions are equal, either exactly or with a degree of freedom correction. This principle can be leveraged to offer a computational solution for cases where \citet{ding-2021}'s results do not hold, i.e. estimates and other information from partial regressions are sufficient to compute correct covariance matrices of coefficient vectors in the multiple regression. If only a subset of coefficients is of interest then estimation and proper statistical inference can be conducted by working with the partial regression alone. 
	
	The current paper is most closely related to \citet{giles-1984} and \citet{ding-2021}. I extend \citet{giles-1984} and and \citet{ding-2021} in several ways. First, while \citet{giles-1984} had demonstrated the YFWL theorem for instrumental variables estimation of an exactly identified model, I extend the result to over-identified models. I offer a general proof that nests both cases of instrumental variables estimation, and of course least squares estimation (because, in this case, the regressors become their own instruments). Second, \citet{giles-1984} had not investigated properties of the \textit{estimated} covariance matrices from full and partial models. I draw on \citet{ding-2021} to do so, and in so doing I also extend the analysis in \citet{ding-2021} from ordinary least squares estimation to generalized linear instrumental variables estimation. Third, neither \citet{giles-1984} nor \citet{ding-2021} had investigated whether the YFWL theorem can be applied to the K-class of estimators, including the LIML estimator, or linear GMM estimators. I do so and thereby extend the coverage of the YFWL theorem significantly.  
	
	I would like to end this introductory section by noting a limitation of the YFWL for generalized instrumental variables estimation. The key results of the theorem, i.e. equality of the coefficient vectors and residual vectors from full and partial models can only be derived when then conditioning set (the set of covariates used for residualizing) does not contain any endogenous variables. The reason has to do with the necessity of projection matrices in deriving key results of the YFWL theorem. If endogenous variables appear in the conditioning set, the relevant projection matrices are no longer available for the analysis. But this negative result is less restrictive than appears at first sight, as I discuss in section~\ref{sec:yfwl-iv-summary}.
	
	The rest of the paper is organized as follows: in section~\ref{sec:setup}, I introduce the set-up and pose the main question; in section~\ref{yfwl-iv}, I present the YFWL for generalized instrumental variables estimation; in section~\ref{sec:kclass}, I discuss why the YFWL theorem applies, in large samples, to all consistent K-class estimators (including the LIML estimator) but does not apply to linear GMM estimators (including the two step GMM estimator); in section~\ref{sec:history}, I present a narrative account of the analytical development of the theorem through the decades; I conclude in section~\ref{sec:conclusion} by highlighting the usefulness of, and highlighting a puzzle about the intellectual history of, the YFWL theorem. Proofs are collected together in the appendix.  
	
	\section{The set-up and the question}\label{sec:setup}
	
	Consider a linear regression of an outcome variable, $Y$, on a set of $k$ covariates, 
	\begin{equation}\label{model-full}
		Y = W \beta + \varepsilon,
	\end{equation}
	where $Y$ is a $N \times 1$ vector of the outcome variable, $W$ is a $N \times k$ matrix of covariates, $\beta$ is a $k \times 1$ of parameters, and $\varepsilon$ is a $N \times 1$ vector of the stochastic error term. 
	
	Suppose on the basis of a specific question under investigation, it is possible for the researcher to partition the set of regressors into two groups, $W_1$ and $W_2$, where the former is a $N \times k_1$ matrix and the latter is a $N \times k_2$ matrix, with $k=k_1+k_2$, i.e. 
	\[
	W = \left[ W_1 : W_2\right]. 
	\]
	The model in (\ref{model-full}) can then be written as 
	\begin{equation}\label{model-full-iv}
		Y = W_1 \beta_1 + W_2 \beta_2 + \varepsilon,
	\end{equation}
	where $\beta_1$ and $\beta_2$ are $k_1 \times 1$ and $k_2 \times 1$ vectors of parameters, so that $\beta' = \left(\beta_1', \beta_2' \right) $. I would like to ensure that the model satisfies the identification condition \citep[p.~19, 223]{greene} for both sets of parameters, $\beta_1$ and $\beta_2$, using the following assumption.
	
	\begin{assumption}\label{reg-col-rank}
		The matrices $W_1$ and $W_2$ are both of full column rank.
	\end{assumption}
	
	Suppose the researcher is only interested in the second group of regressors, $W_2$, i.e. she is only interested in estimating and conducting inference about $\beta_2$. But, of course, she does not want to completely throw away the information in $W_1$ because the variables in this subset do impact $Y$ and are likely to also be correlated to the variables in $W_2$. Hence, while she wants to condition her analysis on $W_1$, she is not interested in their partial effects on the outcome (e.g. fixed effects in a panel data regression model). 
	
	Since the researcher is only interested in the parameters on $W_2$, she could consider the following \textit{partial} regression model,
	\begin{equation}\label{model-partial-iv}
		\widetilde{Y} = \widetilde{W}_2 \tilde{\beta}_2 + \tilde{\varepsilon},
	\end{equation}
	where $\widetilde{Y}$ and $\widetilde{W}_2$ are residualized (or partialled out) versions of the outcome variable and the variables of interest in (\ref{model-full-iv}), respectively, where residualization has been carried out with respect to $W_1$ (the exogenous covariates in the model), i.e. 
	\begin{equation}\label{b-iv}
		\widetilde{Y} = M_{W_1} Y, \text{ and } \widetilde{W}_2 =M_{W_1} W_2,
	\end{equation}
	where
	\begin{equation}
		M_{W_1} = I - P_{W_1} \text{ and } P_{W_1} = W_1 \left( W'_1 W_1\right)^{-1} W'_1,
	\end{equation}
	are the hat and residual maker matrices, respectively, for the set of exogenous regressors, $W_1$. 
	
	The matrices $P_{W_1}$ and $M_{W_1}$ play important roles in the whole analysis. They are both symmetric and idempotent. They are referred to, respectively, as the `hat matrix' and the `residual maker matrix' because $P_{W_1}$ projects any vector in $\mathbb{R}^N$ onto the column space of $W_1$, and $M_{W_1}$ projects the same vector onto the orthogonal complement of the column space of $W_1$. Thus, $P_{W_1}$ generates the vector of predicted values and $M_{W_2}$ generates the vector of least squares residuals for a regression of any vector in $\mathbb{R}^N$ on $W_1$. 
	
	The researcher faces an additional challenge: while the conditioning variables in the model, $W_1$, are exogenous, i.e. $\E \left(\varepsilon | W_1 \right) = 0$, the variables of interest, $W_2$, are endogenous, i.e. $\E \left(\varepsilon | W_2 \right) \neq 0$. Since the model has endogenous regressors, the parameters cannot be estimated consistently by the method of  least squares. Instead, the researcher turns to instrumental variables estimation using a set of $k_3$ instrumental variables, represented by the $n \times k_3$ matrix, $Z_2$, with $k_3 \geq k_2$.
	\begin{assumption}\label{assmp:iv}
		The set of instruments, $Z_2$, satisfy the following conditions \citep[p.~223--224]{greene}:
		\begin{enumerate}
			\item Exogeneity: The instrumental variables are uncorrelated with the error term in the full model, i.e. $\Cov \left( \varepsilon, Z_2\right) = 0$;
			
			\item Relevance: The instrumental variables are correlated with the regressors. In particular, $Z'_2 W_1$ has rank  $ k_1 $, and $ Z'_2 W_2 $ has rank  $ k_2 $.\footnote{Note that both the exogeneity and relevance conditions can also be stated in asymptotic terms using probability limits.}
		\end{enumerate}
	\end{assumption}
	
	Given her interest and the set-up, can the researcher avoid estimating the full model in (\ref{model-full-iv})? Can she work with the partial model (\ref{model-partial-iv}) and yet consistently estimate and conduct proper statistical inference on $\beta_2$ in (\ref{model-full-iv})? The YFWL theorem provides an answer in the affirmative.

	\section{The YFWL theorem for linear instrumental variables estimation}\label{yfwl-iv}
	
	\subsection{Relationship between estimated coefficients}
	The full set of instrumental variables is given by
	\[
	Z = \left[ W_1 : Z_2\right],
	\]
	so that the number of instrumental variables, $k_1+k_3$, can be larger than the number of endogenous variables, $k_1+k_2$, i.e. we are possibly dealing with an overidentified model. Hence, I will consider the 2SLS estimator, also known as the generalized instrumental variables estimator \citep[p.~321]{davidson-mackinnon-2004}, for the full regression model (\ref{model-full-iv}).\footnote{The 2SLS estimator uses, from among all possible linear combination of the set of instruments, the one that minimizes the asymptotic covariance matrix of the instrumental variables estimator \citep{brundy-jorgenson-1971}.} The unique 2SLS estimator for coefficients in the full regression model (\ref{model-full-iv}) is given by \citep[p.~231]{greene}, 
	\begin{equation}\label{bfull-2sls-all}
		b = 
		\begin{pmatrix}
			\underset{\left( k_1 \times 1\right)}{b_1}  \\
			\underset{\left( k_2 \times 1\right)}{b_2} 
		\end{pmatrix} =
		\left[ \left( P_Z W\right) ' W\right]^{-1} \left( P_Z W\right) ' Y,
	\end{equation}
	where $P_Z = Z \left( Z'Z\right)^{-1}Z' $ projects on to the column space of $Z$. Hence, 
	\begin{equation}\label{bfull-2sls}
		b_2 = \text{ ($2,1$) block of } \left[ \left( P_Z W\right) ' W\right]^{-1} \left( P_Z W\right) ' Y.
	\end{equation}
	Turning to the partial model (\ref{model-partial-iv}), we estimate it with the instrumental variables formed by the  residualized set of instrumental variables, $\widetilde{Z}_2=M_{W_1}Z_2$, which is a $N \times k_3$ matrix. Thus, the 2SLS estimator for the partial regression model (\ref{model-partial-iv}) is given by
	\begin{equation}\label{bpartial-2sls}
		\tilde{b}_2 = \left[ \left( P_{\widetilde{Z}_2} \widetilde{W}_2 \right)' \widetilde{W}_2 \right]^{-1} \left( P_{\widetilde{Z}_2} \widetilde{W}_2 \right)' \widetilde{Y},
	\end{equation}
	where $P_{\widetilde{Z}_2} = \widetilde{Z}_2 \left( \widetilde{Z}_2' \widetilde{Z}_2\right)^{-1} \widetilde{Z}_2' $ projects on to the column space of $\widetilde{Z}_2$.
	
	\begin{theorem}\label{thm:coef-resid}
		For instrumental variables estimation of the coefficients in the full regression model (\ref{model-full-iv}) and the partial regression model (\ref{model-partial-iv}), we have
		\begin{enumerate}
			\item the estimated coefficient vectors $b_2$ in (\ref{bfull-2sls}) and $\tilde{b}_2$ in (\ref{bpartial-2sls}) are identical, and
			\item the residual vectors from the full model (\ref{model-full-iv}) and from the partial model (\ref{model-partial-iv}) are identical.
		\end{enumerate}
	\end{theorem}
	
	The results in theorem~\ref{thm:coef-resid} applies to the following special cases: (a) the instrumental variables (IV) estimator for an exactly identified model, when the number of instruments is exactly equal to the number of endogenous variables, i.e. $k_3=k_2$; (b) the OLS estimator, when both sets of regressors in (\ref{model-full}), $W_1$ and $W_2$, are exogenous. The latter case was proved by \citet{udny-1907, frisch-waugh-1933, lovell-1963}; the former case was proved by \citet{giles-1984}. Theorem~\ref{thm:coef-resid} generalizes both.

	\subsection{Relationship between estimated covariance matrices of coefficients}
	
	\subsubsection{Estimated covariance matrices are equal up to a degree of freedom correction}\label{cov-iv}
	Using the result in theorem 9.4 in \citet[p.~264]{greene}, we see that the \textit{estimator} of the asymptotic covariance matrix of $b'=\left( b'_1, b'_2\right) $, the instrumental variables estimator of $\left( \beta'_1, \beta'_2\right) $ in (\ref{model-full}) from the full regression model in (\ref{model-full-iv}), is given by
	\begin{align}
		\widehat{\Var} \begin{pmatrix}
			b_1 \\
			b_2
		\end{pmatrix} & =
		\begin{pmatrix}
			\widehat{\Var} \left( b_1\right)  & \widehat{\E} \left( b_1-\beta_1, b'_2-\beta'_2 \right) \\
			\widehat{\E} \left( b_2-\beta_2, b'_1-\beta'_1 \right) & \widehat{\Var} \left( b_2\right)  
		\end{pmatrix} \nonumber \\ 
		& = 
		\left[ \left(P_Z W\right)' \left(P_Z W\right) \right]^{-1} \left(P_Z W\right)' \widehat{\Omega}_f \left(P_Z W\right) \left[ \left(P_Z W\right)' \left(P_Z W\right) \right]^{-1},
	\end{align}
	where $W = \left[ W_1 : W_2\right]$, $Z = \left[ W_1 : Z_2\right]$, $P_Z = Z\left(Z'Z \right)^{-1} Z'$ and $\widehat{\Omega}_f$ (subscript `f' for identifying the full regression) is an estimate of the covariance matrix of the error term, $\varepsilon$, in the full regression model (\ref{model-full-iv}). Hence, the estimated variance of $b_2$ is given by 
	\begin{equation}\label{estvar-model-full}
		\widehat{\Var} \left( b_2\right) = \text{ (2,2) block of }\left[ \left(P_Z W\right)' \left(P_Z W\right) \right]^{-1} \left(P_Z W\right)' \widehat{\Omega}_f \left(P_Z W\right) \left[ \left(P_Z W\right)' \left(P_Z W\right) \right]^{-1}.
	\end{equation}
	
	Using a similar argument, we see that the estimator of the covariance matrix of $\tilde{b}_2$, the instrumental variables estimator of $\tilde{\beta}_2$ in the partial regression model (\ref{model-partial-iv}), is given by
	\begin{equation}\label{estvar-model-partial}
		\widehat{\Var} \left( \tilde{b}_2\right) = \left[{\left(P_{\widetilde{Z}_2} \widetilde{W}_2\right)}^{'} \left(P_{\widetilde{Z}_2} \widetilde{W}_2\right)\right]^{-1} \left(P_{\widetilde{Z}_2} \widetilde{W}_2\right)' \widehat{\Omega}_p \left(P_{\widetilde{Z}_2} \widetilde{W}_2\right) \left[{\left(P_{\widetilde{Z}_2} \widetilde{W}_2\right)}^{'} \left(P_{\widetilde{Z}_2} \widetilde{W}_2\right)\right]^{-1},
	\end{equation}
	where $\widehat{\Omega}_p$ (subscript `p' for identifying the partial regression) is an estimate of the covariance matrix of the error term, $\tilde{\varepsilon}$, in the partial regression model (\ref{model-partial-iv}).
	
	\begin{theorem}\label{thm:cov}
		If the estimators of the error vectors in the full and partial regression models are equal, i.e.  $\widehat{\Omega}_f=\widehat{\Omega}_p$  (after making degrees of freedom corrections, if necessary), then $\widehat{\Var} \left( b_2\right)$ in (\ref{estvar-model-full}) and $\widehat{\Var} \left( \tilde{b}_2\right)$ in (\ref{estvar-model-partial}) are equal (up to a degrees of freedom correction, if necessary). 
	\end{theorem}
	
	Let me summarize what we get from theorem~\ref{thm:cov}: (a) if the estimate of the error covariance matrix depends only on the vector of regression residuals, then (\ref{estvar-model-full}) and (\ref{estvar-model-partial}) are exactly equal; (b) if some degree of freedom correction is applied to generate the estimate of the error covariance matrix, then (\ref{estvar-model-full}) and (\ref{estvar-model-partial}) are equal up to the relevant degree of freedom correction; (c) if the estimate of the error covariance matrix depends, in addition to the residual vector, on the elements of the matrix of regressors, then (\ref{estvar-model-full}) and (\ref{estvar-model-partial}) are not, in general, equal even after making degree of freedom adjustments. In particular, theorem~\ref{thm:cov} in this paper extends \citet[Theorem~2, 3, 4]{ding-2021} to linear instrumental variables estimation and we have the following results: 
	\begin{enumerate}
		\item In models with homoskedastic errors, the covariance matrices from the multiple and partial regressions are equal up to a degree of freedom adjustment , i.e. $(N-k_2)\widehat{\Var} \left( \tilde{b}_2\right) = (N-k) \widehat{\Var} \left( b_2\right)$. This is because $\widehat{\Omega}_p = 1/(N-k_2) \text{diag} \left[ \tilde{u}_i^2\right] $ and $\widehat{\Omega}_f = 1/(N-k) \text{diag} \left[ u_i^2\right] $, where `diag' denotes a $N \times N$ diagonal matrix and $k=k_1+k_2$.
		
		\item In models with HC0 version of HCCM \citep{mackinnon-white-1985}, the covariance matrices from the multiple and partial regressions are exactly equal because $\widehat{\Omega}_p = \text{diag} \left[ \tilde{u}_i^2\right] $ and $\widehat{\Omega}_f = \text{diag} \left[ u_i^2\right] $.
		
		\item In models with HC1 version of HCCM \citep{mackinnon-white-1985}, the covariance matrices from the multiple and partial regressions are equal up to a degree of freedom adjustment because $\widehat{\Omega}_p = N/(N-k_2)\text{diag} \left[ \tilde{u}_i^2\right] $ and $\widehat{\Omega}_f = N/(N-k) \text{diag} \left[ u_i^2\right] $. Hence, $(N-k_2)\widehat{\Var} \left( \tilde{b}_2\right) = (N-k) \widehat{\Var} \left( b_2\right)$.
		
		\item In models using heteroskedasticity and autocorrelation (HAC) consistent covariance matrices \citep{newey-west-1987}, the covariance matrices from the multiple and partial regressions are exactly equal because $\widehat{\Omega}_p = \left( \omega_{|i-j|} \tilde{u}_i \tilde{u}'_j\right)_{1 \leq i, j, \leq N}  $ and $\widehat{\Omega}_f = \left( \omega_{|i-j|} \tilde{u}_i \tilde{u}'_j\right)_{1 \leq i, j, \leq N} $, as long as the same weights are used in both regressions.
		
		\item In models using the cluster robust estimate of the variance matrix (CRVE) \citep{mackinnon-etal-2023}, the covariance matrices from the multiple and partial regressions are exactly equal if no degree of freedom correction is used or if $G/(G-1)$ is used as the degree of freedom correction, where $G$ is the number of clusters \citep[p.~8]{cameron-miller-2015}. If $G(N-1)/(G-1)(N-k)$ is used as the degree of freedom correction, then we have what \citet{mackinnon-etal-2023} call the $ CV_1 $ feasible CRVE and for this CRVE, the following holds: $(N-k_2)\widehat{\Var} \left( \tilde{b}_2\right) = (N-k) \widehat{\Var} \left( b_2\right)$. 
	\end{enumerate}
	
	In all these cases, researchers can use estimated covariance matrices from the partial regression (\ref{model-partial-iv}), with the relevant degree of freedom adjustment if necessary, to conduct proper statistical inference on the parameters from the multiple regression (\ref{model-full-iv}).
	
	\subsubsection{The cases that were left out}\label{sec:left-out}
	
	There are some cases where the estimate of the error covariance matrix depends, in addition to the residual vector, on elements of the covariate and/or instrumental variables matrix. In these cases, (\ref{estvar-model-full}) and (\ref{estvar-model-partial}) will not be equal, even after making degrees of freedom corrections. Some common cases where this happens are: (a) models using HC2, HC3, HC4, or HC5 forms of HCCM;\footnote{HC2 and HC3 were  introduced by \citet{mackinnon-white-1985}; HC4 was introduced by \citet{cribari-neto-2004}; HC5 was introduced by \citet{cribari-neto-etal-2007}. For a discussion of all the variants of HCCM, see \citet{mackinnon-2013}.} (b) some variants of the HAC covariance matrix and the $CV_2$ and $CV_3$ forms of feasible CRVEs \citep{mackinnon-etal-2023}. In these cases, it is not possible to use standard errors from the partial regression, with or without degrees of freedom correction, for inference about the coefficient vector in the full regression model. But, it is still possible to compute the correct covariance matrix for the coefficient vector in the full regression by using information that becomes available while estimating the partial regression. Thus, estimation of the full regression model can still be avoided.\footnote{See appendix~\ref{app:computational}.}

	\subsection{Limitation of the YFWL theorem for instrumental variables estimation}\label{sec:limitation}
	
	In the YFWL for instrumental variables estimation, I have assumed that the set of variables that are of interest are endogenous and the set of variables used for conditioning are exogenous. Can we switch the role of the endogenous and exogenous variables and still get the YFWL theorem? I would like to show that the answer is in the negative but then also argue that this negative result is not too restrictive. 
	
	Let us return to the full regression model (\ref{model-full-iv}) and specify that the set of variables of interest, $W_2$, is exogenous, and the set of variables used for conditioning (or residualizing), $W_1$, is endogenous. Suppose, now the researcher has a set of $k_3$ valid instrumental variables, $Z_1$, satisfying the exogeneity and relevance conditions, where $k_3 \geq k_1$ . The full set of instrumental variables is $Z = \left[Z_1 : W_2 \right] $, and the instrumental variables estimator of the parameter vector ($\beta'_1, \beta'_2$) is now given by ($b'_1, b'_2$), where,
	\begin{equation}
		\begin{pmatrix}
			b_1 \\
			b_2
		\end{pmatrix} =
		\begin{pmatrix}
			Z'_1 W_1 & Z'_1 W_2 \\
			W'_2 W_1 & W'_2 W_2 
		\end{pmatrix}^{-1}
		\begin{pmatrix}
			Z'_1 Y \\
			W'_2 Y
		\end{pmatrix},
	\end{equation}
	so that, using results for the inverse of partitioned matrices \citep[p.~994]{greene}, we have,
	\begin{equation}\label{bnew-full}
		b_2 = \left[ W'_2 \left( I - W_1 \left(Z'_1 W_1 \right)^{-1}Z'_1 \right)W_2  \right]^{-1} W'_2 \left( I - W_1 \left(Z'_1 W_1 \right)^{-1}Z'_1 \right)Y.  
	\end{equation}
	As can be seen from the above expression, we do not get any projection matrices. Neither can the inverse on the left be opened up because $W_2$ is not a square matrix. 
	
	Now let us turn to the partial regression, which, in this case will be 
	\begin{equation}\label{model-partial-iv-new}
		\widetilde{Y} = \widetilde{W}_2 \tilde{b}_2 + \tilde{u},
	\end{equation}
	where $\widetilde{Y} = M_{W_1}Y$ and $\widetilde{W}_2 = M_{W_1} W_2$. Since $\widetilde{W}_2$ is orthogonal to $W_1$ (the endogenous variables), it is itself exogenous to the error of the full regression model. Hence, (\ref{model-partial-iv-new}) can be estimated by the method of ordinary least squares, giving us
	\begin{equation}\label{bnew-part-ols}
		\tilde{b}_{2, OLS} = \left( {\widetilde{W}_2}' \widetilde{W}_2\right)^{-1} {\widetilde{W}_2}' \widetilde{Y}.  
	\end{equation}
	If, instead instrumental variables estimation is conducted with the residualized instrumental variables, $\widetilde{Z}_1 = M_{W_1} Z_1 = Z_1$ (because $Z_1$ is orthogonal to $W_1$), we will have
	\begin{equation}\label{bnew-part-iv}
		\tilde{b}_{2, IV} = \left( Z'_1 \widetilde{W}_2\right)^{-1} Z'_1 \widetilde{Y}.  
	\end{equation}
	In general, neither (\ref{bnew-part-ols}) nor (\ref{bnew-part-iv}) is equal to (\ref{bnew-full}). Hence, the neat results of the YFWL theorem cannot be derived in this case, i.e. for instrumental variables estimation when the set of conditioning variables are endogenous. This negative result is less restrictive than might appear, as I discuss below by way of summarizing the YFWL for instrumental variables estimation.
	
	\subsection{Summary}\label{sec:yfwl-iv-summary}
	
	Let me summarize the facts about the YFWL theorem and highlight when it is applicable in terms of properties of the regressors and how they can be partitioned into two subsets. In the case of least squares estimation, ordinary or generalized, the YFWL theorem is applicable without any restrictions, i.e. the set of regressors can be partitioned into two subsets any which way. For instrumental variables estimation, there are some restrictions that need to be kept in mind. These restrictions arise because two different distinctions are at play that might not coincide: exogenous variables versus endogenous variables; and variables of interest versus those not of direct interest (and hence relegated to the conditioning set). 
	
	Let $S_W$, $S_{X}$ and $S_{I}$ denote, respectively, the full set of covariates, the subset of exogenous variables and the subset of variables of interest. Let $S_{N}=S_W \setminus S_{X}$ denote the set of endogenous variables, and let $S_{D}=S_W \setminus S_{I}$ denote the set of conditioning variables (variables not of direct interest but necessary for conditioning). 
	\begin{enumerate}
		\item When $S_I = S_N$, then the YFWL theorem is applicable for instrumental variables estimation (using the residualized set of instrumental variables) as shown in theorem~\ref{thm:coef-resid} and theorem~\ref{thm:cov}. 
		
		\item If $S_I \supset S_N$, then the YFWL theorem is applicable for instrumental variables estimation. Since the conditioning set does not contain any endogenous variables, the results of theorem~\ref{thm:coef-resid} and theorem~\ref{thm:cov} apply. The only change that is required is to expand $W_1$ (the set of endogenous variables) to include the exogenous variables that are of interest and to then use these variables as their own instrumental variables.
		
		\item If $S_I \subset S_N$, then the YFWL theorem can be made applicable for instrumental variables estimation in the following way: take the set of variables of interest as being equal to the set of endogenous variables (or, equivalently, take the conditioning set to be the set of all the exogenous variables only) and apply the result of case~1 (so that results of theorem~\ref{thm:coef-resid} and theorem~\ref{thm:cov} apply); finally, extract coefficients of interest and their covariance matrix.
		
		\item If $S_I \supseteq S_X$, then the the YFWL theorem for instrumental variables estimation cannot be applied, as demonstrated in section~\ref{sec:limitation} (for the case when $S_I = S_X$ or equivalently $S_D = S_N$). Note that this is the only case when the YFWL theorem is not applicable for instrumental variables estimation cannot be applied. Hence, the negative result is less restrictive than it might have appeared at first sight. After all, it only restricts one out of four possibilities 
		
	\end{enumerate}

	\section{Does the FWL theorem apply to K-class and linear GMM estimators?}\label{sec:kclass}
	
	The LIML estimator (a member of the K-class estimators) and the linear GMM estimator are closely related to the 2SLS estimator (another member of the K-class of estimators) discussed above. Hence, it is natural to ask whether the FWL theorem, which I have derived above for the 2SLS estimator, applies to the K-class and linear GMM estimators.\footnote{The popular \texttt{ivreg2} suite of functions in STATA seems to allow for partialling for all these estimators. The paper that explains the technical details of \texttt{ivreg2} claims in an informal argument, by relying on the FWL theorem, that: ``The invariance of the estimation results to partialling-out applies to one- and two-step estimators such as OLS, IV, LIML, and two-step GMM, but not to CUE or to GMM iterated more than two steps.'' \citep[p.~484]{baum-etal-2007}. The paper does not offer any formal proofs of these claims.} I will demonstrate that: (a) the YFWL theorem applies, in large samples, to all K-class estimators for which $K \to 1$, and (b) the YFWL theorem applies to the 2SGMM estimator but does not apply to general linear GMM estimators. The clue to arriving at this answer is offered by trying to understand the common structure of the 2SLS, K-class and linear GMM estimators, and then using a crucial step from the proof of theorem~\ref{thm:coef-resid}.\footnote{For a discussion of the K-class of estimators, see \citet[p.~649]{davidson-mackinnon}.} 
	
	\subsection{The theory}
	To get started, let us recall from (\ref{bfull-2sls}) that the 2SLS estimator of $b_2$ in the full model (\ref{model-full-iv}) is given by
	\begin{equation}\label{b-full-2sls-1}
		\text{ the (2,1) block of }\left[ W' P_Z W \right]^{-1} W' P_ZY, 
	\end{equation}
	and the 2SLS estimator of $\tilde{b}_2$ in the partial model (\ref{model-partial-iv}) is given by
	\begin{equation}\label{b-partial-2sls-1}
		\left[ \widetilde{W}'_2 P_{\widetilde{Z}_2} \widetilde{W}_2\right]^{-1} \widetilde{W}'_2 P_{\widetilde{Z}_2} Y.
	\end{equation}	 
	In the proof of theorem~\ref{thm:coef-resid}, I showed that (\ref{b-full-2sls-1}) and (\ref{b-partial-2sls-1}) were equal. What is important is to note that the proof of theorem~\ref{thm:coef-resid} relies crucially on $P_Z$ in (\ref{b-full-2sls-1}) being a projection matrix, i.e. a symmetric and idempotent $n \times n$ matrix \citep[p.~164]{strang-2006}. This is important because when $P_Z$ is a projection matrix, we can decompose it into two orthogonal matrices, $P_{W_1}$ and $P_{\widetilde{Z}_2}$ \citep[p.~323]{rao-etal-2008}. It is because of this decomposition that I could show the crucial equality in theorem~\ref{thm:coef-resid}. This means that if I can express any estimator in the same form as the 2SLS estimator in (\ref{bfull-2sls-all}), then I can replicate the argument that established theorem~\ref{thm:coef-resid}. This will then allow me to claim that the YFWL theorem applies to that estimator. This will be my general strategy to investigate whether the YFWL theorem applies to the K-class and linear GMM estimators.
	
	\subsubsection{K-class of Estimators}
	
	\textit{General case.} The K-class of estimators for $b$ in the full model (\ref{model-full-iv}) is given by \citep[p.~649]{davidson-mackinnon}:
	\begin{equation}\label{kclass}
		b = \left[ W' \left(I - K M_Z \right) W \right]^{-1} W' \left(I - K M_Z \right) Y,
	\end{equation}
	where $K$ is a scalar and $M_Z = I - Z\left(Z'Z \right)Z'$ is the residual maker matrix with respect to the set of instrumental variables. When $K=0$, we get the OLS estimator; when $K=1$, we get the 2SLS estimator; when $K=\hat{\kappa}$, where $\hat{\kappa}$ is the smallest eigenvalue of $\left(U'M_{Z}U \right)^{-1/2} U'M_{Z}U \left(U'M_{Z}U \right)^{-1/2}$, where $U=\left[ Y : W_2\right] $, we get the LIML estimator.\footnote{Note that $\left(W'M_{Z}W \right)^{-1/2}$ denotes the inverse of the square root factor of the positive definite matrix $U'M_{Z}U$.} 
	
	With regard to the K-class of estimators in (\ref{kclass}), note that, in general, $\left(I - K M_Z \right)$ is symmetric, because $M_Z$ is symmetric, but not idempotent. Hence, $\left(I - K M_Z \right)$ is not a projection matrix for a general value of $K$. This implies that, for a general value of $ K $, the K-class of estimators does not have the same structure as the 2SLS estimator in (\ref{bfull-2sls-all}). Thus, the arguments that establish theorem~\ref{thm:coef-resid} and ~\ref{thm:cov} will not work. Hence, my conjecture is that for a general value of $K$, the YFWL does not apply to the K-class of estimators. 
	
	\textit{Special case.} But there is an important subset of K-class estimators where it does. To identify this subset, note that for all K-class estimators, where $K$ tends to $1$ at an appropriately fast rate, $\left(I - K M_Z \right)$ is approximately idempotent (in large samples) because  
	\[
	\left(I - KM_Z \right)\left(I - KM_Z \right) = I - KM_Z + \left( K^2 - K\right)M_Z \approx I - M_Z
	\]
	where I have used the idempotency of $M_Z$ and the fact that $K \approx 1$ in large samples.\footnote{K-class estimators, where $K$ tends to $1$ at a rate faster than $n^{-1/2}$, are consistent \citep[p.~649]{davidson-mackinnon}.} Thus, for K-class estimators where $K$ tends to $1$---which includes 2SLS, LIML, Fuller, but not OLS---$\left(I - KM_Z \right)$ behaves approximately like the projection matrix $P_Z=\left(I - M_Z \right)$ in large samples. Thus, these K-class estimators have the same structure as the 2SLS estimator in (\ref{bfull-2sls-all}) and the arguments of theorem~\ref{thm:coef-resid} and ~\ref{thm:cov} will apply. That is why the YFWL theorem \textit{can be applied approximately in large samples} for the subset of K-class estimators for which $K \to 1$ with sample size, including the LIML estimator.\footnote{An alternative argument might be as follows: since the LIML is the limit of an iterative process involving OLS estimators, and since the YFWL theorem applies to OLS estimators, the YFWL theorem is also applicable to the LIML estimator. This argument is incomplete unless one shows that the set of estimators that form the sequence in the iterative process belong to a closed set. If it is an open set, then the limit (LIML) will not necessarily inherit the properties of the estimators that are sequences in the iterative process.}

	\subsubsection{Linear GMM Estimators}
	\textit{General case.} The linear GMM estimator for $b$ in the full model (\ref{model-full-iv}) is given by \cite[p.~185]{cameron-trivedi-2005}:
	\begin{equation}\label{gmm}
		b = \left[ W' Z W_N Z' W \right]^{-1} W' Z W_N Z' Y,
	\end{equation}
	where $W_N$ is some symmetric, positive definite $k_3 \times k_3$ weighting matrix (which is necessarily of full rank). Note that, in general, while $ZW_NZ'$ is symmetric, it is not idempotent.\footnote{As a counterexample, consider $W_N=I$ (the identity matrix of the correct size), which is symmetric and of full rank. Hence, $ZW_NZ'=ZZ'$. Now note that, in general, $ZZ'ZZ'$ is not equal to $ZZ'$. Thus, $ZW_NZ'$ is not idempotent. Hence, it cannot be a projection matrix. For purposes of illustrating the weighting matrix, \citet[p.~173]{cameron-trivedi-2005} also use $W_N=I$.} This implies that in the linear GMM estimator in (\ref{gmm}), $ZW_NZ'$ is not a projection matrix, even in large samples and not even approximately. Hence, the arguments that establishes equality of the coefficient vectors from the full and partial regressions in theorem~\ref{thm:coef-resid} will not work. Hence, my conjecture is that the FWL theorem cannot be applied to general linear GMM estimators.\footnote{This negative result for linear GMM estimators pertain to overidentified models. For exactly identified models, the linear GMM estimator reduces to the IV estimator, for which we can apply the FWL theorem, as I have shown in theorem~\ref{thm:coef-resid} and theorem~\ref{thm:cov}.}
	
	\textit{Special case.} But there is a special case of the linear GMM estimator where the YFWL theorem does apply. This is the 2SGMM estimator that uses the specific weighting matrix, $W_N$, to minimize the asymptotic variance of the estimator. For $b$ in the full model (\ref{model-full-iv}) the 2SGMM arises with the choice of $W_N = N\left( Z'DZ\right)^{-1} $, where $D$ is a $N \times N$ diagonal matrix with the principal diagonal populated by $\widehat{u}_i^2$, where $\widehat{u}$ is the $N \times 1$ vector of residuals of the full model estimated by some consistent estimator, e.g. 2SLS \cite[p.~187]{cameron-trivedi-2005}. Thus, the two step linear GMM (2SGMM) estimator for $b$ is given by
	\begin{equation}\label{2sgmm}
		b = \left[ W' Z \left( Z'DZ\right)^{-1} Z' W \right]^{-1} W' Z \left( Z'DZ\right)^{-1} Z' Y,
	\end{equation}
	where the $N$ in the definition of $W_N$ has canceled out. 
	
	Let $\widehat{Z} = D^{1/2}Z$, where $D^{1/2}$ is the $N \times N$ matrix with the principal diagonal populated by $\widehat{u}$ (residuals from the full model). Now define $P_{\widehat{Z}} = D^{1/2}\widehat{Z} \left( \widehat{Z}' D \widehat{Z} \right)^{-1} \widehat{Z}' D^{1/2} = D^{1/2} \widehat{Z} \left( \widehat{Z}' D^{1/2} D^{1/2} \widehat{Z} \right)^{-1} \widehat{Z}' D^{1/2}$. It is easy to see that $P_{\widehat{Z}}'=P_{\widehat{Z}}$ and
	\[
	 P_{\widehat{Z}} P_{\widehat{Z}} = D^{1/2}\widehat{Z} \left( \widehat{Z}' D \widehat{Z} \right)^{-1} \widehat{Z}' D^{1/2} D^{1/2}\widehat{Z} \left( \widehat{Z}' D \widehat{Z} \right)^{-1} \widehat{Z}' D^{1/2} = D^{1/2}\widehat{Z} \left( \widehat{Z}' D \widehat{Z} \right)^{-1} \widehat{Z}' D^{1/2} = P_{\widehat{Z}}.
	\]
	Thus, $P_{\widehat{Z}}$ is a symmetric and idempotent matrix. Hence, $P_{\widehat{Z}}$ is a projection matrix \citep[p.~164]{strang-2006}. Let $\widehat{Y} = D^{-1/2}Y$, $\widehat{W} = D^{-1/2}W$, where $D^{-1/2}$ is the $N \times N$ matrix with the principal diagonal populated by the reciprocal of $\widehat{u}$. The two step linear GMM estimator for $b$ in (\ref{2sgmm}) can be written as 
	\begin{equation}\label{2sgmm-1}
		b = \left[ \left( P_{\widehat{Z}}\widehat{W}\right)' \widehat{W} \right]^{-1} \left( P_{\widehat{Z}} \widehat{W}\right)' \widehat{Y}.
	\end{equation}
	Thus, the two step linear GMM estimator in (\ref{2sgmm-1}) has the same structure as the 2SLS estimator in (\ref{bfull-2sls-all}), including, crucially, the role played by a projection matrix in both. Hence, the arguments of theorem~\ref{thm:coef-resid} and ~\ref{thm:cov} can be applied to (\ref{2sgmm-1}). This implies that the YFWL applies to the 2SGMM estimator.

	\subsection{An empirical example}
	As an empirical example, I return to David Card's $ 1995 $ study of returns to schooling using geographical proximity as instrumental variable \citep{card-1995}. The model is
	\begin{equation}\label{card-1995}
		\log y = \beta_0 + \beta_1 * \text{SCHOOLING} + X \gamma + u
	\end{equation}
	where $y$ is earnings (wage in cents per hour in $1976$), $SCHOOLING$ is years of schooling, and $X$ is a set of exogenous controls like experience (and its square), race (whether black), whether lived in the South, whether lived in an SMSA, and regional dummy variables. 
	
	The model (\ref{card-1995}) is estimated with OLS, 2SLS, LIML, Fuller's K-class estimator, IGMM (a GMM estimator where the identity matrix of the correct dimension is used as the weighting matrix) and 2SGMM. Other than OLS, estimation treats $SCHOOLING$ as endogenous and uses two instrumental variables to deal with endogeneity: (a) proximity to $2$ year college; (b) proximity to $4$ year college.\footnote{I use \texttt{ivmodel} in R to get the data and implement the K-class of estimators. I write my own code for the IGMM and 2SGMM estimators. For the IGMM estimator, I use an identity matrix of the correct size as the weighting matrix. For the 2SGMM estimator, I implement the following steps: In the first step, I estimate the model by 2SLS. Using the residuals from the first stage, I construct the optimal weighting matrix. In the second step, I use the optimal weighting matrix (from the first step) and solve the linear system of equations whose solution appears as the 2SGMM estimator in (\ref{gmm}). R code for this example is available here: \url{https://drive.google.com/file/d/1ug0SZVwuW5vuxBHZbsVQWlaNXoxxNBMW/view?usp=sharing}.} The source of the data is the National Longitudinal Survey of Young Men (NLSYM) and the sample size is $3101$.\footnote{``The NLSYM began in 1966 with 5525 men age 14--24 and continued with follow-up surveys through 1981.'' \citep{card-1995}.} 
	
	Two sets of results are presented in Table~\ref{table:card}. Row~1 reports results of estimating the full model; row~2 presents estimates of the partial model, where the latter model partials out the full set of exogenous variables from the outcome variable (log earnings), from the endogenous variable (years of schooling) and from the two instrumental variables (proximity to $2$ year and $4$ year college). From Table~\ref{table:card}, we see that the parameter estimates from the full and partial models are identical for the K-class of estimators (OLS, 2SLS, Fuller and LIML), and also for the 2SGMM estimator.\footnote{In a previous version of this paper, an error in my R code had given rise to different 2SGMM estimates for the full and partial model. I would like to thank Kit Baum for pointing out that my 2SGMM results were possibly incorrect.} But the estimates are different between the full and partial model for the IGMM estimator: $0.13753$ for the full model and $0.16274$ for the partial model. This confirms the theoretical argument presented in the previous sub-section as to why the FWL applies to the common K-class of estimators and to the 2SGMM estimator, but not to a general linear GMM estimator. 
	
	\begin{table}[hbt!]
		\begin{center}
			\begin{threeparttable}
				\caption{Estimates of Returns to Schooling from Full and Partial Models}
				\label{table:card}
				\begin{tabular}{lcccccc}
					\toprule
					& OLS & 2SLS & Fuller & LIML & IGMM & 2SGMM \\ 
					\hline
					Full model & 0.07469 & 0.15706 & 0.15826 & 0.16403 & 0.13753 & 0.15521 \\ 
					Partial model & 0.07469 & 0.15706 & 0.15828 & 0.16403 & 0.16274 & 0.15521 \\  
					
					\bottomrule
				\end{tabular}
				\begin{tablenotes}
					\small
					\item \textit{Notes:} The table presents estimates of the return to schooling in a regression of log earnings on schooling, where two instrumental variables are used for estimation: proximity to $ 2 $ year college; proximity to $ 4 $ year college. The partial model partials out \textit{all} the exogenous variables from the outcome, endogenous and instrumental variables. OLS = ordinary least squares; 2SLS = two stage least squares; Fuller = Fuller's modified LIML; LIML = limited information maximum likelihood; IGMM = GMM with the identity matrix as the weighting matrix; 2SGMM = two step optimal GMM.  
				\end{tablenotes}
			\end{threeparttable}
		\end{center}
		
	\end{table}	
	
	\section{Historical evolution of the YFWL theorem}\label{sec:history}
	
	\subsection{Pioneering contribution of G. Udny Yule}\label{sec:yule}
	
	\subsubsection{Novel notation}
	
	\citet{udny-1907} considers the relationship among $n$ random variables $X_1$, $ X_2$, $\ldots$, and $X_n$ using a sample which has $N$ observations on each of the variables. In particular, suppose this relationship is captured by a linear regression of the first variable, $X_1$, on the other variables, $X_2, \ldots, X_n$, and a constant. This linear regression is what Yule called a \textit{multiple regression}. With reference to (\ref{model-full}), therefore \citet{udny-1907} uses: $k=n$, $X_1=Y$, $X_2$ the second column of $W$, $X_3$ the third column of $W$, and so on.
	
	Subtracting the sample mean from all variables, the multiple regression equation can be equivalently expressed in `deviation from mean' form in which the constant drops out. Let $x_1, x_2, \ldots, x_n$ denote the same random variables, but now expressed as deviations from their respective sample means. Introducing a novel notation to express the coefficients and the residual, \citet{udny-1907} writes the sample analogue of the multiple regression function as follows:
	\begin{equation}\label{lrg-1}
		x_1 = \underbrace{b_{12.34 \ldots n} x_2 + b_{13.24 \ldots n} x_3 + \cdots + b_{1n.234 \ldots n-1} x_n}_{\text{predicted value}} + \underbrace{x_{1.234 \ldots n}}_{\text{residual}}.
	\end{equation}	 
	
	In (\ref{lrg-1}), the subscripts of the coefficient on each regressor, as also the residual, is divided into the primary and secondary subscripts. The primary subscripts come before the period; the secondary subscripts come after the period. For the coefficients of regressors, the primary subscripts identify the dependent variable and that particular regressor, in that order; and the secondary subscripts identify the other regressors in the model. For the residual in (\ref{lrg-1}), $x_{1.234 \ldots n}$, the primary subscript identifies the dependent variable and the secondary subscripts identify all the regressors.
	
	It is useful to note three features of the subscripts. First, the order of the primary subscripts is important but the order among the secondary ones are not because the order in which the regressors are arranged in immaterial for estimation and inference. For instance, $b_{12.34 \ldots n}$ denotes the coefficient on $x_2$ for a regression of $x_1$ on $x_2, x_3, \ldots, x_n$. Similarly, $b_{1j.234 \ldots n}$ denotes the coefficient on $x_j$ for a regression of $x_1$ on $x_2, x_3, \ldots, x_n$, and note that the secondary subscripts excludes $ j $. Second, elements of the subscripts cannot be repeated because it is not meaningful to include the dependent variable as a regressor or repeat a regressor itself. Third, since the coefficients of the multiple regression function are related to partial correlations, the notation was also used for denoting relevant partial correlations too.\footnote{See, for instance, equation (2) in \citet{udny-1907}.}

	
	\subsubsection{The theorem}
	
	Now consider, with reference to the multiple regression (\ref{lrg-1}), what Yule called \textit{partial regressions}. These refer to regressions with partialled out variables. For instance, with reference to (\ref{lrg-1}), the partial regression of $x_1$ on $x_2$ would be constructed as follows: (a) first run a regression of $x_1$ on all the variables other than $x_2$, i.e. $x_3, x_4, \ldots, x_n$, and collect the residuals; (b) run a regression of $ x_2 $ on all the other regressors, i.e. $x_3, x_4, \ldots, x_n$, and collect the residuals; (c) run a regression of the residuals from the first step (partialled out outcome variable) on the residuals from the second step (partialled out first regressor). 
	
	\citet[p.~184]{udny-1907} showed that, if parameters of the multiple regression and the partial regression(s) were estimated by the method of least squares then the corresponding coefficients would be numerically the same. For instance, considering the first regressor, $x_2$, in (\ref{lrg-1}), he demonstrated that
	\begin{equation}\label{yule-1}
		\frac{\sum \left( x_{1.34 \ldots n} \times x_{2.34 \ldots n}\right) }{\sum \left( x_{2.34 \ldots n}\right)^2 } = b_{12.34 \ldots n}  
	\end{equation}
	where the summation runs over all observations in the sample.\footnote{For a proof see appendix~\ref{app:yule}.}
	
	To see why (\ref{yule-1}) gives the claimed result, recall that $x_{1.34 \ldots n}$ is the residual from a regression of $ x_1 $ on $ x_3, \ldots, x_n $ and $x_{2.34 \ldots n}$ is the residual from the regression of $ x_2 $ on $ x_3, \ldots, x_n $. By construction, both have zero mean in the sample. Recall, further, that in a regression of a zero-mean random $y$ on another zero-mean random variable $z$, the coefficient on the latter is $\sum y z /\sum z^2$. Hence, the left hand side of (\ref{yule-1}) is the coefficient in a regression of $x_{1.34 \ldots n}$ on $x_{2.34 \ldots n}$. The right hand side is, of course, the coefficient on $x_2$ in (\ref{lrg-1}). Hence, the result.
	
	The above argument, of course, can be applied to any of the regressors in (\ref{lrg-1}) and thus \citet{udny-1907} proved the following general result: the coefficient on any regressor in (\ref{lrg-1}) is the same as the coefficient in a bivariate regression of the residualized dependent variable (the vector of residual from a regression of the dependent variable on the other regressors) and the residualized regressor (the vector of residuals from a regression of the regressor of interest on the other regressors). 
	
	With reference to the question posed in section~\ref{sec:setup}, \citet{udny-1907}, therefore, showed that a researcher could work with partial regressions if she were only interested in a subset of the coefficients in the multiple regression. She would arrive at the same estimate of the parameters by estimating partial regressions as she would if she had estimated the full model. In particular, when $W_1$ contained only one variable (other than the column of $1$s), \citet{udny-1907} provided the following answer: (a) run a regression of $y$ (demeaned $Y$) on the variables in $w_2$ (demeaned variables in $W_2$), and collect the residuals; (b) run a regression of $w_1$ (demeaned variable in $W_1$, i.e. excluding the constant) on the variables in $w_2$, and collect the residuals; (c) regress the first set of residuals on the second to get the desired coefficient.
	
	There are two things to note about Yule's answer. First, he did not provide an answer to the question posed in section~\ref{sec:setup} when $W_1$ contained more than one random variable (excluding the constant). Of course, partial regressions could be estimated for each independent variable, but they had to be separately estimated as bivariate regressions with \textit{all} the other variables used for the partial regressions. Second, \citet{udny-1907} did not investigate the relationship among variances of the parameters estimated from multiple and partial regressions. Is the \textit{estimated} standard error of a coefficient also identical from the full and partial regressions? Yule did not pose, and hence did not provide any answers to, this question. 
	
	\subsubsection{Application to a regression with a linear time trend}
	
	We can immediately apply the result from Yule's theorem given above to a regression with a linear time trend. With (\ref{lrg-1}), let $x_n$ denote the demeaned linear time trend variable. Applying (\ref{yule-1}), we will have the following result: the coefficient on any of the regressors, $x_2, x_3, x_{n-1}$ in (\ref{lrg-1}) is the same as the coefficient in a bivariate regression of residualized $x_1$ on the relevant regressor. This is of course the exact same result that was presented $26$ later in \citet{frisch-waugh-1933}. 
	
	\subsection{Frisch and Waugh's result}\label{sec:frischwaugh}
	
	\subsubsection{Notation and result}
	\citet[p.~389]{frisch-waugh-1933} study the relationship among $n+1$ variables, one of which is a linear time trend. In reference to (\ref{model-full}), therefore \citet{frisch-waugh-1933} use: $k=n$, $X_0=Y$, $X_1$ the second column of $W$, $X_2$ the third column of $W$, and so on. They use the same convention of considering variables in the form of deviations from their respective sample means.\footnote{``Consider now the $ n $ variables $x_0, \ldots, x_{n-1}$ and let time be an $ (n+1)$th variable $x_n$. Let all the variables be measured from their means so that $\sum x_i=0 \quad (i = 0, \ldots, n)$ where $\sum$ denotes a summation over all the observations.'' \citep[p.~394]{frisch-waugh-1933}.} Using the \textit{exact same notation} as used by \citet{udny-1907} and postulating an \textit{a priori} true linear relationship among these variables, they write,
	\begin{equation}\label{fw-1}
		x_0 = \beta_{01.34 \ldots n} x_1 + \beta_{02.24 \ldots n} x_2 + \cdots + \beta_{0n.234 \ldots n-1} x_n,
	\end{equation}
	and consider estimating the parameters of (\ref{fw-1}) in two ways. First, they consider the multiple regression of $x_0$ on  $x_1, \ldots, x_n$:
	\begin{equation}\label{fw-2}
		x_0 = b_{01.34 \ldots n} x_1 + b_{02.24 \ldots n} x_2 + \cdots + b_{0n.234 \ldots n-1} x_n.
	\end{equation}
	Second, they consider the multiple regression of $x'_0$ on  $x'_1, \ldots, x'_{n-1}$ (note that $x_n$ has been excluded from the set of regressors),
	\begin{equation}\label{fw-3}
		x'_0 = b'_{01.34 \ldots n} x'_1 + b'_{02.24 \ldots n} x'_2 + \cdots + b'_{0,n-1.234 \ldots n-1} x'_{n-1},
	\end{equation}
	where the primed variables are the corresponding time-demeaned original variables, i.e., for $j = 0, 1, \ldots, n-1$, $x'_j$ is the residual in the regression of $x_j$ on $x_n$.\footnote{Note that (\ref{fw-1}), (\ref{fw-2}) and (\ref{fw-3}) are just the predicted regressions functions. These equations exclude the residuals.} Using the basic rules of determinants and Cramer's rule for solving linear equation systems, \citet[p.~394--396]{frisch-waugh-1933} demonstrated that the coefficients denoted by $b$ in (\ref{fw-2}) are numerically equal to the corresponding coefficients denoted by $b'$ in (\ref{fw-3}).\footnote{For a proof see appendix~\ref{app:frisch}. In his presentation of the Frisch-Waugh theorem, \citet[p.~84--86]{chipman-1998} argues as if Frisch and Waugh had used projection matrices in their proof. That is not correct. \citet{frisch-waugh-1933} did not use projection matrices in their proof.} 
	
	With regard to the question posed in section~\ref{sec:setup}, Frisch and Waugh provide the same answer as Yule: coefficients are the same whether they are estimated from the multiple or from partial regressions. There are both similarities and differences between \citet{udny-1907}  and \citet{frisch-waugh-1933}. First, whereas in \citet{udny-1907}, only one variable could be included in $W_1$ (the subset of covariates that was of interest to the researcher), in \citet{frisch-waugh-1933} only one random variable could be included in  $W_2$ (the subset of covariates that was \textit{not} of interest to the researcher). Second, much like \citet{udny-1907} before them, \citet{frisch-waugh-1933} did not investigate the relationship among estimated variances of the parameters of multiple and partial regressions. The question that is relevant for statistical inference, i.e. standard errors, had still not been posed.

	\subsection{Lovell extends the OLS analysis}\label{sec:lovell}
	
	\citet{lovell-1963} extended the reach of the theorem significantly and addressed both questions that had been left unanswered by \citet{udny-1907} and \citet{frisch-waugh-1933}. On the one hand, \citet{lovell-1963} partitioned the set of regressors, $W$, into two subsets without any restrictions on the number of variables in each subset; on the other, he laid the groundwork for thinking about the \textit{estimated} covariance matrices of the coefficient vectors.
	
	In the context of OLS estimation, \citet{lovell-1963} demonstrated two important results: (a) the coefficient vectors are numerically the same whether they are estimated from the multiple or the partial regressions, and (b) the vector of residuals from the multiple and partial regressions are numerically the same.\footnote{I omit the proofs because they are just special cases of theorem~\ref{thm:cov} in this paper.} The first result completed the YFWL so far as the estimate of the coefficient is concerned because the partitioning of the set of regressors was completely general; the second result laid the groundwork for comparing estimated variances of the coefficient vectors from multiple and partial regressions.\footnote{It is straightforward to extend the YFWL from ordinary least squares to generalized least squares estimation, as Lovell had noted when commenting on autocorrelated errors \citep[p.~1004]{lovell-1963}. Other scholars have worked on variations of Lovell's results; see, for instance, \citet{fiebig-bartels-1996, krishnakumar-2006}.}
	
	\subsection{D. Giles extends the theorem to IV estimation}
	
	So far, the YFWL theorem was always posed in the context of least squares, ordinary or generalized, estimation. To the best of my knowledge, \citet{giles-1984} was the first scholar to extend the theorem to instrumental variables estimation of an exactly identified model, i.e. in the setting where the number of instruments was exactly equal to the number of endogenous variables.\footnote{``In the following discussion the necessary inverse matrices are assumed to exist, and we consider only the case of equal numbers of regressors and instruments as this includes such common estimators as two-stage least squares and other members of the k-class.'' \citep[p.~340]{giles-1984}. The last part of the sentence uses non-standard terminology. In current terminology, the two-stage least squares estimator refers to the case where the number of instruments is larger than the number of endogenous variables.} \citet{giles-1984} demonstrated that the YFWL theorem holds for the IV estimator (instrumental variables estimator when the model is exactly identified) under $13$ combinations of partialling. He shows that the coefficient vectors from the full and partial regression models are identical and that their asymptotic covariance matrices are the same. \citet{giles-1984} did not comment on the \textit{estimators} of asymptotic covariance matrices of the coefficient vector and, therefore, did not discuss the need for degrees of freedom corrections.
	
	\subsection{P. Ding considers standard errors for OLS estimation}\label{sec:ding}
	
	For more than $100$ years, the evolution of the theorem had primarily focused on comparing the coefficient vectors from multiple and partial regressions. The YFWL theorem had established that they are numerically the same. This was a very useful result but still did not answer the question that would be relevant if a researcher were interested in conducting statistical inference about the coefficient vector. To conduct inference about $b_2$ using results of estimating $\tilde{b}_2$, we also need to be able to compare their \textit{estimated} covariance matrices. While \citet{lovell-1963} had commented briefly on this issue, to the best of my knowledge, \citet[Theorem~2, 3, 4]{ding-2021} provided the first systematic treatment. \citet{ding-2021} demonstrated that, in the context of ordinary least squares estimation, covariance matrices of coefficient vectors from partial and full regressions would, in many contexts, be exactly equal or equal up to a degrees of freedom correction. I have extended these sets of results to instrumental variables estimation and have commented on the different cases in section~\ref{cov-iv} above.	
	
	\section{Discussion and conclusion}\label{sec:conclusion}
	
	By way of concluding this paper, I would like to offer two sets of comments. My first set of comments is about the substance and usefulness of the YFWL theorem, and the second set is about a puzzle in the history of the YFWL theorem. 
	
	There are two reasons that make the YFWL theorem so useful, the first computational and the second conceptual. Let me start with the computational usefulness. For a linear regression model estimated by the method of least squares, the YFWL theorem shows that if a researchers is interested in estimating only a subset of the parameters of the model, she need not estimate the full model. She can partition the set of regressors into two subsets, those that are of interest and those that are not of direct interest (but needs to be used for conditioning). She can regress the outcome variable on the conditioning set to create the residualized outcome variable. She can regress each of the variables of interest on the conditioning set to create corresponding residualized covariates. Finally, she can regress the residualized outcome variable on the residualized covariates of interest to get the estimated parameter of interest. 
	
	If, in addition, the variables of interest are endogenous and she has a set of valid instrumental variables, she can carry out the same procedure as above for instrumental variable estimation. In addition to creating the residualized outcome variable and the residualized covariates of interest (the endogenous variables), she must create residualized instrumental variables with respect to the exogenous variables in the model. Now, if she estimates an instrumental variables regression of the residualized outcome variable on the residualized covariates of interest using the residualized instrumental variables, then she will get consistent estimates of the parameters of interest. 
	
	In large samples, the same procedure can be justified for the subset of K-class estimators (LIML being an important member of this class) for which $K \to 1$ with sample size because the YFWL theorem applies approximately to this class of estimators in large samples. It is important to note that the same procedure cannot be used for linear GMM estimators in general because the YFWL does not apply to such estimators. But the YFWL does apply to the 2SGMM estimator in linear models. This means that the YFWL theorem applies to OLS estimators, linear instrumental variables estimators, all K-class estimators for which $K \to 1$ with sample size and the 2SGMM estimator. Thus the results in this paper significantly increases the coverage of the YFWL theorem for linear models.
	
	For statistical inference, she can work with a partial regression in many, but not all, cases. If the errors are homoskedastic, then the estimated covariance matrix from the partial regression can be used for inference once a degree of freedom adjustment is made. If the errors are heteroskedastic, as is often the case in \textit{cross sectional} data sets, and the researcher wishes to use the HC0 or HC1 variant of HCCM, then she can use the estimated covariance matrix from the partial regression without any adjustment in the case of HC0 and with a degree of freedom adjustment in the case of HC1. If the researcher is using a \textit{time series} data set and wishes to use a HAC covariance matrix, then she can use the estimated covariance matrix from the partial regression as is. If the researcher is using a \textit{panel data} set and wishes to use the standard cluster robust covariance matrix, then she can use the estimated covariance matrix from the partial regression without any adjustment; if a degree of freedom correction is used to compute the cluster robust covariance matrix, then she will need to apply a corresponding degree of freedom adjustment. 
	
	If, on the other hand, the researcher wishes to use the HC2, HC3, HC4 or HC5 variants of heteroskedasticity consistent covariance matrix, or if she wishes to use other variants of the cluster robust covariance matrix or HAC covariance matrices, then she cannot use above results, i.e. the covariance matrices from multiple and partial regressions are no longer equal, even after degrees of freedom adjustments. Instead, the researcher can use the four-step computational method proposed in appendix~\ref{app:computational} if she wishes to use information from the partial regression to conduct proper statistical inference about parameters in the multiple regression.
	
	While the question of computation is certainly important, the YFWL theorem is perhaps even more relevant because of its indispensability for an intuitive understanding of multiple regression and partial correlations. In non-experimental settings, whenever a researcher has to analyze the relationship among three of more variables, she has to address a question that G. Udny Yule, a pioneer of regression analysis, raised more than a century ago.  
	\begin{quote}
		[F]or instance, it might be found that changes in pauperism were highly correlated (positively) with changes in the out-relief ratio, and also with changes in the proportion of old; and the question might arise how far the first correlation was due merely to a tendency to give out-relief more freely to the old than the young, i.e. to a correlation between changes in out-relief and changes in the proportion of the old. \citep[p.~225]{yule-1911}.
	\end{quote}
	In other words, how would a researcher distinguish between the true, or partial, correlation between changes in pauperism and out-relief ratio from the multiple (or total) correlation that was being observed between these two variables merely because both were correlated with a third variable, the proportion of the old. 
	
	To address this question in the most general setting, \citet{yule-1911} goes on to explain, the researcher must posit a linear relationship between the $n$ variables of interest, $X_1, X_2, X_3, \ldots, X_n$. Such a relationship, Yule, notes 
	\begin{quote}
		... will be of the form
		\[
		X_1 = a + b_2 X_2 + b_3 X_3 + \cdots + b_n X_n.
		\]
		If in such a generalized \textit{regression} or \textit{characteristic equation} we find a sensible positive value for any one coefficient such as $b_2$, we know that there must be a positive correlation between $X_1$ and $X_2$ that cannot be accounted for by mere correlations of $X_1$ and $X_2$ with $X_3, X_4$, or $X_n$ for the effects of changes in these variables are allowed for in the remaining terms on the right. The magnitude of $b_2$ gives, in fact, the mean change in $X_1$ associated with a unit change in $X_2$ when all the remaining variables are kept constant. \citep[p.~226, emphasis in the original]{yule-1911}.
	\end{quote}
	But how do we know that the ``magnitude of $b_2$ gives, in fact, the mean change in $X_1$ associated with a unit change in $X_2$ when all the remaining variables are kept constant?'' We know this precisely because of the YFWL theorem, which has rigorously established that $b_2$ is the correlation coefficient (normalized by the ratio of their standard deviations) between residualized version of $X_1$ and $X_2$. By residualizing both $X_1$ and $X_2$ with respect to $X_3, X_4, \ldots, X_n$, we have removed from the picture the possible influence these other variables might have had on $X_1$ and $X_2$. Thus, we are now able to measure the correlation between $X_1$ and $X_2$ after removing the possible influence of the other variables out of the way, i.e. holding ``all the remaining variables ... constant.'' This essential conceptual meaning of `holding variables constant' in a multiple regression rests completely on the YFWL theorem.\footnote{Yule was a leading figure behind the development of regression analysis. The substantive issue he seems to have analyzed in many studies is the effect of the out-relief ratio (proportion of poor provided monetary, food or other assistance but not required to be in an institution, compared to those who were required to enter an institution like a poor house to receive support) on pauperism. In \citet{yule-1899}, he demonstrated that out-relief is positive correlated with, perhaps causes and increase in, pauperism. Of course, in substantive terms, Yule's regression analysis suffers from simultaneity bias. After all pauperism might also be a cause of administrative measures that change the out-relief ratio.}
	
	My second comment relates to a puzzle in the intellectual history of the YFWL theorem. The puzzle arises from noting that  \citet{udny-1907}'s result is essentially the same as the result proved by \citet{frisch-waugh-1933}, as I have demonstrated. What is puzzling, therefore, is that \citet{frisch-waugh-1933} do not cite \citet{udny-1907}. This omission seems puzzling to me given two fact. First, \citet{frisch-waugh-1933} refer to a previous paper by one of the authors: \citet{frisch-1929}. It is also to be noted that Frisch had published a related work in 1931 \citep{frisch-mudgett-1931}. What is interesting is that in both these papers, there is explicit reference to the partial correlation formula introduced by Yule.\footnote{For instance, see  \citet[p.~75, 76]{frisch-1929} and \citet[p.~389]{frisch-mudgett-1931}} Recall that Yule had introduced a novel notation for representing partial correlations and coefficients in a multiple regression equation in his 1907 paper that I have discussed. The notation was then introduced to a wider readership in statistics with his 1911 book \citep{yule-1911}. This book was extremely popular and ran several editions, later ones with Kendall \citep{yule-kendall-1948}. Second, \citet{frisch-waugh-1933} use the exact same notation to represent the regression function that Yule had introduced in his 1907 paper. as I have highlighted. 
	
	These two facts suggest that Ragnar Frisch was familiar with Yule's work. In fact, Yule's work on partial correlations and multiple regression was the standard approach that was widely used and accepted by statisticians in the early part of the twentieth century \citep{griffin-1931}.  Therefore, it is a puzzle of intellectual history as to why \citet{frisch-waugh-1933} did not cite Yule's $ 1907 $ result about the equality of multiple and partial regression coefficients which was, in substantive terms, exactly what they went on to prove in their paper. Of course, \citet{frisch-waugh-1933} used a different method of proof from \citet{udny-1907}, as I have demonstrated in this paper. But in substantive terms, \citet{frisch-waugh-1933} proved the same result that Yule had established more than two decades ago. No matter what the reason for the omission in history, it seems that now is the right time to acknowledge Yule's pioneering contribution---both in terms of computation and in terms of interpretation---by attaching his name to a theorem that is so widely used in econometrics.
	
	\bibliographystyle{apalike}
	\bibliography{yfwl_refs}
	
	\begin{appendices}
		
		\section{Proof of Theorem~\ref{thm:coef-resid}}\label{app:yfwl}
		
		\subsection{Coefficients are identical}
		\begin{proof}
			
			Let us first consider the full model (\ref{model-full-iv}) and note that the 2SLS estimator of the coefficient vector is given by
			\begin{equation}
				b = 
				\begin{pmatrix}
					\underset{\left( k_1 \times 1 \right) }{b_1}  \\
					\underset{\left( k_1 \times 1 \right) }{b_2} 
				\end{pmatrix} =
				\left[ \left( P_Z W\right)' \left( P_Z W \right) \right]^{-1} \left( P_Z W\right)' Y  = \begin{pmatrix}
					\underset{\left( k_1 \times n \right) }{A_{11}} \\
					\underset{\left( k_2 \times n \right) }{A_{21}}
				\end{pmatrix} Y,
			\end{equation}
			so that					
			\begin{equation}\label{b2-alt-1}
				b_2 = A_{21}Y.
			\end{equation}
			
			Now consider the partial model (\ref{model-partial-iv}), and note that the 2SLS estimator of the coefficient vector is given by 
			\begin{equation}\label{b2-alt-2}
				\tilde{b}_2 = \left[ \left( P_{\widetilde{Z}_2} \widetilde{W}_2 \right)' \widetilde{W}_2 \right]^{-1} \left( P_{\widetilde{Z}_2} \widetilde{W}_2 \right)' \widetilde{Y} = \left[ \left( P_{\widetilde{Z}_2} \widetilde{W}_2 \right)' \widetilde{W}_2 \right]^{-1} \left( P_{\widetilde{Z}_2} \widetilde{W}_2 \right)' Y 
			\end{equation}
			because
			\[
			\left( P_{\widetilde{Z}_2} \widetilde{W}_2 \right)' \widetilde{Y} = \widetilde{W}'_2 P_{\widetilde{Z}_2} M_{W_1} Y = \widetilde{W}'_2 P_{\widetilde{Z}_2}  Y = \left( P_{\widetilde{Z}_2} \widetilde{W}_2 \right)' Y   
			\]
			where, the second equality follows because $P_{\widetilde{Z}_2} M_{W_1} = P_{\widetilde{Z}_2} $ by the idempotency of $M_{W_1}$.\footnote{$P_{\widetilde{Z}_2} M_{W_1} = \widetilde{Z}_2 \left( \widetilde{Z}'_2 \widetilde{Z}_2\right)^{-1} \widetilde{Z}'_2 M_{W_1} = \widetilde{Z}_2 \left( \widetilde{Z}'_2 \widetilde{Z}_2\right)^{-1} Z'_2 M_{W_1} M_{W_1} = \widetilde{Z}_2 \left( \widetilde{Z}'_2 \widetilde{Z}_2\right)^{-1} Z'_2 M_{W_1} = \widetilde{Z}_2 \left( \widetilde{Z}'_2 \widetilde{Z}_2\right)^{-1} \widetilde{Z}'_2 = P_{\widetilde{Z}_2}$.}
			
			Comparing (\ref{b2-alt-1}) and the far right hand side of (\ref{b2-alt-2}), we see that the proof will be complete, i.e. $b_2 = \tilde{b}_2$ will be established, if I can now show that
			\begin{equation}\label{proof-key}
				A_{21} = \left[ \left( P_{\widetilde{Z}_2} \widetilde{W}_2 \right)' \left( P_{\widetilde{Z}_2} \widetilde{W}_2 \right) \right]^{-1} \left( P_{\widetilde{Z}_2} \widetilde{W}_2 \right)'.
			\end{equation}
			
			We know from the argument in \citet[p.~6]{ding-2021}, that if $X = \left[ X_1 : X_2 \right]$, then 
			\[
			\left( X' X\right)^{-1} X' =   
			\begin{pmatrix}
				* \\
				\left( {\widetilde{X}_2}^{'} \widetilde{X}_2\right)^{-1} \widetilde{X}_2^{'},
			\end{pmatrix}
			\] 
			where $\widetilde{X}_2 = \left[  I - X_1 \left( X'_1 X_1\right)^{-1} X'_1\right] X_2 $, and the ($1,1$) block on the RHS, $*$, is left unspecified because it is not needed for the computations. Now, let $X_1 = P_Z W_1$ and $X_2 = P_Z W_2$, so that $P_Z W = \left[P_Z W_1 : P_Z Z_2\right] $. I will demonstrate that $\widetilde{X}_2 =P_Z \widetilde{W}_2= P_{\widetilde{Z}_2} \widetilde{W}_2$ in two steps and that will complete the proof. 
			
			\textit{First step:} Since $Z = [W_1:Z_2]$, Lemma~2 in \citet[p.~5]{ding-2021} shows that $P_Z = P_{W_1}+P_{\widetilde{Z}_2}$ and $P_{W_1} P_{\widetilde{Z}_2} = 0$.\footnote{This result on the decomposition of projection matrices is used in \citet[p.~85]{chipman-1998} and \citet[page~323]{rao-etal-2008}.} Hence, $P_Z W_1 = P_{W_1}W_1+P_{\widetilde{Z}_2} W_1 = P_{W_1}W_1 = W_1$, because the first term is zero,
			\[
			P_{\widetilde{Z}_2} W_1 = \widetilde{Z}_2\left({\widetilde{Z}_2}' \widetilde{Z}_2 \right)^{-1} {\widetilde{Z}_2}' W_1 = \widetilde{Z}_2\left({\widetilde{Z}_2}' \widetilde{Z}_2 \right)^{-1} Z'_2 \left[  I - W_1 \left( W'_1 W_1\right)^{-1} W'_1\right] W_1 = 0,
			\]
			and the second term is
			\[
			P_{W_1}W_1 = \left[ W_1 \left( W'_1 W_1\right)^{-1} W'_1 \right] W_1 = W_1. 
			\] 
			Hence,
			\begin{align*}
				\widetilde{X}_2 & = \left[  I - P_Z W_1 \left( W'_1 P_Z W_1\right)^{-1} W'_1 P_Z\right] P_Z W_2  \\
				& = \left[  P_Z - P_Z W_1 \left( W'_1 P_Z W_1\right)^{-1} W'_1 P_Z P_Z \right]  W_2 \\
				& = P_Z \left[  I - W_1 \left( W'_1 P_Z W_1\right)^{-1} W'_1 P_Z \right]  W_2  \\
				& = P_Z \left[  I - W_1 \left( W'_1 W_1\right)^{-1} W'_1 \right]  W_2 \\
				& = P_Z M_{W_1} W_2 \\
				& = P_Z \widetilde{W}_2, 
			\end{align*}
			where I have used the idempotency of $P_Z$ and the fact that $P_Z W_1 = W_1$, which was shown above. 
			
			\textit{Second step:} Note that  
			\[
			P_Z \widetilde{W}_2 = \left( P_{W_1}+P_{\widetilde{Z}_2}\right) \widetilde{W}_2 = P_{W_1} \widetilde{W}_2 + P_{\widetilde{Z}_2} \widetilde{W}_2 = P_{W_1} M_{W_1} W_2 + P_{\widetilde{Z}_2} \widetilde{W}_2 = 0 + P_{\widetilde{Z}_2} \widetilde{W}_2,
			\]
			because $P_{W_1} \perp M_{W_1}$.
			
			\subsection{Residuals are identical}
			Let $u_M$ and $u_P$ denote the vectors of residuals from the instrumental variables estimation of the multiple and partial regression models (\ref{model-full-iv}) and (\ref{model-partial-iv}), respectively. Note two properties of the residual maker matrix $M_{W_1}$: first, that $M_{W_1}u_M = \left( I - W_1 \left(W'_1 W_1 \right)^{-1} W'_1\right) u_M = u_M$ because $W'_1 u_M=0$ (by the exogeneity of $W_1$); and, second, that $M_{W_1}W_1= \left( I - W_1 \left(W'_1 W_1 \right)^{-1} W'_1\right) W_1=0$ (because the vector of residuals from projecting each column of $W_1$ onto the column space of $W_1$ is the vector of zeros). 
			
			Since $b_2 = \tilde{b}_2$, we can see immediately that residuals from the instrumental variables estimation of the multiple and partial regression models are the same because,
			\begin{equation}\label{resid-equal}
				u_M = M_{W_1}u_M = M_{W_1}\left( Y - W_1 b_1 - W_2 b_2\right) =  \widetilde{Y}  - 0 - \widetilde{W}_2 b_2 = \widetilde{Y} - \widetilde{W}_2 \tilde{b}_2 = u_P,
			\end{equation}
			where I have used $b_2 = \tilde{b}_2$ in the penultimate equality.
		\end{proof}

		\section{Proof of Theorem~\ref{thm:cov}}\label{app:cov}
		
		\begin{proof}
			Comparing (\ref{estvar-model-full}) and (\ref{estvar-model-partial}), we see that the proof will be completed if I can show that
			\[
			\text{ the (2,1) block of }\left[ \left(P_Z W\right)' \left(P_Z W\right) \right]^{-1} \left(P_Z W\right)' \text{ is equal to } \left[{\left(P_{\widetilde{Z}_2} \widetilde{W}_2\right)}^{'} \left(P_{\widetilde{Z}_2} \widetilde{W}_2\right)\right]^{-1} \left(P_{\widetilde{Z}_2} \widetilde{W}_2\right)'.
			\]
			But this was already shown in the proof of theorem~\ref{thm:coef-resid} when (\ref{proof-key}) was established.
		\end{proof}

		\section{Proof of Yule's result}\label{app:yule}
		
		The first step of the proof is to derive the normal equations arising from the least squares method of estimation. Given the random variables $x_1, x_2, \ldots, x_n$, which are demeaned versions of $X_1, X_2, \ldots, X_n$, the method of least squares chooses constants $b_1, b_2, \ldots, b_n$ to minimize
		\[
		\sum \left( x_1 - b_2 x_2 - b_3 x_3 - \cdots - b_n x_n\right)^2, 
		\]  
		where the sum runs over all observations in the sample. The first order conditions of this minimization problem are referred to as the `normal equations'. Differentiating the above with respect to $b_j$, we have
		\begin{equation}
			\sum x_j \left( x_1 - b_2 x_2 - b_3 x_3 - \cdots - b_n x_n\right) = 0 \quad \left( j=2, 3, \ldots, n\right), 
		\end{equation}
		so that, using Yule's notation, the normal equations are given by
		\begin{equation}\label{yule-norm-eq}
			\sum x_j x_{1.234 \ldots n} = 0 \quad \left( j=2, 3, \ldots, n\right), 
		\end{equation}
		which shows that the residual in the regression equation (\ref{lrg-1}) is uncorrelated with all the regressors included in the model, a result that holds for any regression function whose coefficients are estimated with the method of least squares.
		
		Now consider, $x_{2.34 \ldots n}$, the residuals from a regression of $x_2$ on $x_3, x_4, \ldots, x_n$, and note that it is a linear function of $x_2,x_3, x_4, \ldots, x_n$. Hence, using (\ref{yule-norm-eq}), we have
		\[
		\sum x_{1.234 \ldots n} x_{2.34 \ldots n} = \sum x_{1.234 \ldots n} \left( x_2 - c_1 x_3 - \cdots - c_{n-2} x_n \right)  = 0,
		\]
		for some constants $c_1, \ldots, c_{n-2}$. 
		
		We are now ready to prove the main result: coefficients from multiple and partial regressions are numerically the same.
		\begin{align*}
			0 = & \sum  x_{2.34 \ldots n} x_{1.234 \ldots n} \\
			= &  \sum x_{2.34 \ldots n} \left( x_1 - b_{12.34 \ldots n} x_2 + b_{13.24 \ldots n} x_3 + \cdots + b_{1n.234 \ldots n-1} x_n \right) \\
			= & \sum x_{2.34 \ldots n} \left( x_1 - b_{12.34 \ldots n} x_2\right) \\
			= & \sum x_{2.34 \ldots n} x_1 -  b_{12.34 \ldots n} \sum x_{2.34 \ldots n} x_2 \\
			= & \sum x_{2.34 \ldots n} \left( b_{13.4 \ldots n} x_3 + b_{14.3 \ldots n} x_4 + \cdots + b_{1n.34 \ldots n} x_n + x_{1.34\ldots n}\right) -  b_{12.34 \ldots n} \sum x_{2.34 \ldots n} x_2 \\
			= & \sum x_{2.34 \ldots n} x_{1.34\ldots n} -  b_{12.34 \ldots n} \sum x_{2.34 \ldots n} x_2 \\
			= & \sum x_{2.34 \ldots n} x_{1.34\ldots n} -  b_{12.34 \ldots n} \sum x_{2.34 \ldots n} \left( b_{23.4 \ldots n} x_3 + b_{24.3 \ldots n} x_4 + \cdots + b_{2n.34 \ldots n} x_n + x_{2.34\ldots n}\right) \\
			= & \sum x_{2.34 \ldots n} x_{1.34\ldots n} -  b_{12.34 \ldots n} \sum x_{2.34 \ldots n} x_{2.34\ldots n}
		\end{align*}
		Hence, we have (\ref{yule-1}):
		\[
		\frac{\sum \left( x_{1.34 \ldots n} \times x_{2.34 \ldots n}\right) }{\sum \left( x_{2.34 \ldots n}\right)^2 } = b_{12.34 \ldots n}.
		\]

		\section{Proof of Frisch and Waugh's result}\label{app:frisch}
		
		Consider (\ref{fw-2}) and let
		\begin{equation}\label{def-mij}
			m_{ij} = \sum x_i x_j \quad \left( i,j = 0, 1, 2, \ldots, n\right), 
		\end{equation}
		where the sum runs over all observations, and $x_0, x_1, \ldots, x_n$ are demeaned versions of $X_0, X_1, \ldots, X_n$. Then, \citet[p.~394]{frisch-waugh-1933} assert that the regression equation in (\ref{fw-2}) can be expressed as the following equation that specifies the determinant of the relevant ($n+1$)-dimensional matrix to be zero: 
		\begin{equation}\label{det-ne}
			\begin{vmatrix}
				x_0 & x_1 & \cdots & x_n \\ 
				m_{10} & m_{11} & \cdots & m_{1n} \\ 
				\vdots & \vdots & \ddots & \vdots \\ 
				m_{n0} & m_{n1} & \cdots & m_{nn}  
			\end{vmatrix} = 0.
		\end{equation}
		To see this, which \citet{frisch-waugh-1933} do not explain, perhaps because it was obvious to them, we need to recall two things. First, if we expand the determinant in (\ref{det-ne}) using the first row of the matrix, we will get an equation of the following form,
		\begin{equation}\label{dim:nplus1}
			a_0 x_0 + a_1 x_1 + \cdots + a_n x_n = 0,
		\end{equation}
		where $a_0$ is the determinant obtained by deleting the first row and first column (of the matrix whose determinant is being considered in (\ref{det-ne})), $a_1$ is $-1$ times the determinant obtained by deleting the first row and second column, $a_2$ is $1$ times the determinant obtained by deleting the first row and third column, and so on.\footnote{The signs alternate because the determinant obtained by deleting the first row and the $j$-th column is multiplied by $ -1^{(1+j)}$, where $j=1, 2, \ldots, n+1$. } Assuming $a_0 \neq 0$, which is guaranteed as long as the regressors are not perfectly collinear, this gives
		\begin{equation}\label{dim:nplus1-1}
			x_0  =  -\frac{a_1}{a_0} x_1 - \cdots  -\frac{a_n}{a_0} x_n.
		\end{equation}
		This has the same form as (\ref{fw-2}) and all we need to do is to show that the coefficients in (\ref{fw-2}) are what appears in (\ref{dim:nplus1-1}). To do so, we can use the normal equation and Cramer's rule. 
		
		Recall that the normal equations that had been written in (\ref{yule-norm-eq}) can, with reference to the least squares estimation (\ref{fw-2}), be written more compactly as $X'X b = X'y$, where $X = \left[ x_1 : x_2 : \cdots : x_n\right] $ is the matrix obtained by stacking the regressors column-wise, $y = x_0$ is the dependent variable, and $b$ is the least squares coefficient vector:
		\[
		b = \left[ b_{01.34 \ldots n} \quad b_{02.24 \ldots n} \quad \cdots \quad b_{0n.234 \ldots n-1}\right]. 
		\] 
		Note that the ($i,j$)-th element of $X'X$ is $m_{ij}$ as defined in (\ref{def-mij}), and the $i$-th element of $X'y$ is $m_{i0}$. Thus, the normal equations can be written as 
		\begin{equation}\label{sys-ne}
			\begin{bmatrix}
				m_{11} & m_{12} & \cdots & m_{1j} & \cdots & m_{1n} \\ 
				\vdots & \vdots & \cdots &  \vdots & \cdots  & \vdots \\ 
				m_{n1} & m_{n2} & \cdots & m_{nj} & \cdots & m_{nn}  
			\end{bmatrix} 
			\begin{bmatrix}
				b_{1} \\ 
				\vdots \\ 
				b_{n}  
			\end{bmatrix}
			=  \begin{bmatrix}
				m_{10} \\ 
				\vdots \\ 
				m_{n0}  
			\end{bmatrix}.
		\end{equation}
		Applying Cramer's rule \citep[p.~221]{strang-2006} to this equation system to solve the $b$ vector, keeping track of how many columns are switched and recalling that switching rows (columns) of a matrix only changes the sign of the determinant \citep[p.~203]{strang-2006}, we see that the coefficients in (\ref{fw-2}) and (\ref{dim:nplus1-1}) are identical. Thus, for $j=1,2, \ldots, n$
		\[
		b_j = \frac{\left| B_j \right| }{\left| A\right| },  
		\] 
		where
		\[
		A = \begin{bmatrix}
			m_{11} & m_{12} & \cdots & m_{1j} & \cdots & m_{1n} \\ 
			\vdots & \vdots & \cdots &  \vdots & \cdots  & \vdots \\ 
			m_{n1} & m_{n2} & \cdots & m_{nj} & \cdots & m_{nn}  
		\end{bmatrix} 
		\]
		and 
		\[
		B_j = \begin{bmatrix}
			m_{11} & m_{12} & \cdots & m_{10} & \cdots & m_{1n} \\ 
			\vdots & \vdots & \cdots &  \vdots & \cdots  & \vdots \\ 
			m_{n1} & m_{n2} & \cdots & m_{n0} & \cdots & m_{nn}  
		\end{bmatrix} 
		\]
		is obtained from $A$ by replacing the $j$-th column with $(m_{10} \quad \ldots \quad m_{n0})'$.

		Now consider (\ref{fw-3}) and let 
		\begin{equation}\label{def-mij}
			m'_{ij} = \sum x'_i x'_j \quad \left( i,j = 0, 1, 2, \ldots, n\right), 
		\end{equation}
		where the sum runs, once again, over all observations. Using the same logic as above, we will be able to see that the regression equation in (\ref{fw-3}) can be expressed as
		\begin{equation}\label{det-ne-1}
			\begin{vmatrix}
				x'_0 & x'_1 & \cdots & x'_{n-1} \\ 
				m'_{10} & m'_{11} & \cdots & m'_{1,n-1} \\ 
				\vdots & \vdots & \ddots & \vdots \\ 
				m'_{n-1,0} & m'_{n-1,1} & \cdots & m'_{n-1,n-1}  
			\end{vmatrix} = 0.
		\end{equation}
		
		The strategy will now be to use (\ref{det-ne}) and (\ref{det-ne-1}) to show that the first $n-1$ coefficients in (\ref{fw-2}) are equal to the corresponding coefficients in (\ref{fw-3}). The first thing is to relate $m_{ij}$ and $m'_{ij}$. This is easy to do by noting that 
		\begin{equation}
			x'_j = x_j - \frac{m_{jn}}{m_{nn}}x_n \quad \left( j = 0, 1, 2, \ldots, n-1\right),
		\end{equation} 
		because $x'_j$ is the residual from a regression of $x_j$ on $x_n$. Multiplying $x_i$ on both sides and summing over all observations for $x_i$ and $x_j$, we get
		\begin{equation}\label{mij}
			m'_{ij} = m_{ij} - \frac{m_{in} m_{jn}}{m_{nn}}.
		\end{equation}
		The second step is to return to (\ref{det-ne}) and carry out a series of elementary row operations that converts elements in the second through the penultimate row from $m_{ij}$ to $m'_{ij}$. From the row beginning with $m_{10}$ in (\ref{det-ne}) subtract $-m_{1,n}/m_{n,n}$ times the last row; from the row beginning with $m_{20}$ subtract $-m_{2,n}/m_{n,n}$ times the last row; and so on. Note that these row operations do not change the determinant of the matrix \citep[p.~204]{strang-2006}. Hence, these series of elementary row operations will convert (\ref{det-ne}) to 
		\begin{equation}\label{det-ne-2}
			\begin{vmatrix}
				x_0 & x_1 & \cdots & x_{n-1} & x_n \\ 
				m'_{10} & m'_{11} & \cdots & m'_{1,n-1} & 0 \\ 
				\vdots & \vdots & \ddots & \vdots & \vdots \\ 
				m'_{n-1,0} & m'_{n-1,1} & \cdots & m'_{n-1,n-1} & 0 \\ 
				m_{n0} & m_{n1} & \cdots & m_{n,n-1}& m_{nn}  
			\end{vmatrix} = 0.
		\end{equation}
		because of (\ref{mij}). Now consider the following determinant equation
		\begin{equation}\label{det-ne-3}
			\begin{vmatrix}
				x'_0 & x'_1 & \cdots & x'_{n-1} & 0 \quad \\ 
				m'_{10} & m'_{11} & \cdots & m'_{1,n-1} & 0 \quad \\ 
				\vdots & \vdots & \ddots & \vdots & \vdots \quad \\ 
				m'_{n-1,0} & m'_{n-1,1} & \cdots & m'_{n-1,n-1} & 0 \quad \\ 
				c_0 & c_1 & \cdots & c_{n-1}& 1  \quad
			\end{vmatrix} = 0,
		\end{equation}
		for some arbitrary constants $c_0, \ldots, c_{n-1}$, and by expanding the determinant by the last column, note that (\ref{det-ne-3})  is equivalent to (\ref{det-ne-1}). 
		
		Expanding (\ref{det-ne-2}) by the first row, we get
		\begin{equation}\label{eq:n}
			a_0 x_0 + a_1 x_1 + \cdots + a_{n-1} x_{n-1} + a_n x_n = 0
		\end{equation}
		and expanding (\ref{det-ne-3}) by the first row, we get
		\begin{equation}\label{eq:n-1}
			a'_0 x'_0 + a'_1 x'_1 + \cdots + a'_{n-1} x'_{n-1} = 0,
		\end{equation}
		where
		\begin{equation}\label{eq:coeff}
			a_i = a'_i m_{nn} \quad (i=0,1, 2, \ldots, n-1).
		\end{equation}
		Rearranging (\ref{eq:n}) as
		\begin{equation}
			x_0  =  -\frac{a_1}{a_0} x_1 - \cdots  -\frac{a_{n-1}}{a_0} x_{n-1} -\frac{a_n}{a_0} x_n,
		\end{equation}
		and rearranging (\ref{eq:n-1}) as
		\begin{equation}\label{dim:n-1}
			x'_0  =  -\frac{a'_1}{a'_0} x'_1 - \cdots  -\frac{a'_{n-1}}{a'_0} x'_{n-1},
		\end{equation}
		and noting that (\ref{det-ne-2}) is a re-expression of (\ref{fw-2}) while (\ref{det-ne-3}) is a re-expression of (\ref{fw-3}), we have
		\begin{equation}
			b_{0i.34 \ldots n} = b'_{0i.34 \ldots n} \quad (i=1, 2, \ldots, n-1).
		\end{equation}

		\section{Computational extension of Ding's results}\label{app:computational}
		
		Let me discuss the HC2 case in detail; the other cases can be dealt with in a similar manner. Let $h_{ii}=W'_i\left(W' W\right)^{-1}W_i$, where $W_i$ is the $i$-th column of $W = \left[ W_1 : W_2\right] $ with reference to (\ref{model-full-iv}). The HC2 estimated covariance matrix of $b_2$ in (\ref{model-full-iv}) is given by  
		\begin{equation}\label{hc2-full}
			\text{the ($2,2$) block of }\left(W' W\right)^{-1} W' \frac{\text{diag}\left[ u_i^2\right]}{1-h_{ii}} W \left(W' W\right)^{-1}, 
		\end{equation}
		
		which, using the argument about block matrices in \citet[p.~6]{ding-2021}, is 
		\begin{equation}
			\text{the ($2,2$) block of }
			\begin{pmatrix}
				* \\
				\left( {\widetilde{W}_2}^{'} \widetilde{W}_2\right)^{-1} {\widetilde{W}_2}^{'}
			\end{pmatrix}
			\frac{\text{diag}\left[ u_i^2\right]}{1-h_{ii}} 
			\begin{pmatrix}
				{\widetilde{W}_2} \left( {\widetilde{W}_2}^{'} \widetilde{W}_2\right)^{-1}  & *
			\end{pmatrix}, 
		\end{equation}
		and, therefore, is given by
		\begin{equation}
			\left( {\widetilde{W}_2}^{'} \widetilde{W}_2\right)^{-1} {\widetilde{W}_2}^{'} \frac{\text{diag}\left[ u_i^2\right]}{1-h_{ii}} {\widetilde{W}_2} \left( {\widetilde{W}_2}^{'} \widetilde{W}_2\right)^{-1},
		\end{equation}
		which, since $u_i^2$ can be replaced with $\tilde{u}_i^2$, is
		\begin{equation}\label{hc2}
			\left( {\widetilde{W}_2}^{'} \widetilde{W}_2\right)^{-1} {\widetilde{W}_2}^{'} \frac{\text{diag}\left[ \tilde{u}_i^2\right]}{1-h_{ii}} {\widetilde{W}_2} \left( {\widetilde{W}_2}^{'} \widetilde{W}_2\right)^{-1}.
		\end{equation}
		
		If we can estimate $h_{ii}$ using information available after estimating the partial regression (\ref{model-partial-iv}), we can use (\ref{hc2}) to compute the covariance matrix that is necessary to conduct proper statistical inference on the coefficient vector $b_2$ in the multiple regression (\ref{model-full-iv}). In fact, using results on the inverse of partitioned matrices, we can do so. Hence, the following procedure for estimation and inference coefficient vector $b_1$ in the multiple regression (\ref{model-full-iv}) based on the estimation of the partial regression (\ref{model-partial-iv}) can be suggested.
		
		\begin{enumerate}
			\item Compute $\widetilde{Y} = \left[  I - \widetilde{W}_1 ({\widetilde{W}}'_1 \widetilde{W})^{-1} {\widetilde{W}}' \right]  Y$, and $\widetilde{W}_2 = \left[  I - \widetilde{W}_1 ({\widetilde{W}}'_1 \widetilde{W})^{-1} {\widetilde{W}}' \right] W_2$.
			
			\item Estimate the partial regression (\ref{model-partial-iv}) by regressing $\widetilde{Y}$ on $\widetilde{W}_2$ and get the coefficient vector $\tilde{b}_2$.
			
			\item For $i = 1, 2, , \ldots, N$, compute $h_{ii}=W'_i\left(W' W\right)^{-1}W_i$ for the matrix of covariates, $W = \left[ W_1 : W_2\right]$, in the multiple regression (\ref{model-full-iv}). Let $\widetilde{W}_2 = \left[  I - \widetilde{W}_1 ({\widetilde{W}}'_1 \widetilde{W})^{-1} {\widetilde{W}}' \right] W_2$, and note that, using results for the inverse of partitioned matrices \citep[p.~994]{greene}, we have
			\begin{equation}\label{inv-part}
				\left(W' W\right)^{-1} = \begin{pmatrix}
					W^{11} & W^{12} \\
					W^{21} & W^{22}
				\end{pmatrix},
			\end{equation} 
			where 
			\[
			W^{11} = \left( W'_1 W_1\right)^{-1} + \left( W'_1 W_1\right)^{-1}W'_1 W_2 \left( {\widetilde{W}_2}' \widetilde{W}_2\right)^{-1} W'_2 W_1 \left( W'_1 W_1\right)^{-1},
			\]
			\[
			W^{12}={W^{11}}^{'}= - \left( {\widetilde{W}_2}' \widetilde{W}_2\right)^{-1} W'_2 W_1 \left( W'_1 W_1\right)^{-1}
			\]
			and
			\[
			W^{22} = \left( {\widetilde{W}_2}' \widetilde{W}_2\right)^{-1}.
			\] 
			Thus, we only compute $W^{11}$, $W^{12}$ and $W^{22}$, and then stack them to get $\left( W'W\right)^{-1} $. This avoids computing inverse of the $k \times k$ matrix $W'W$, and instead only involves computing inverses of $k_1 \times k_1$ or $k_2 \times k_2$ matrices.
			
			\item Use the matrix of covariates, $\widetilde{W}_2$ and the vector of residuals, $\tilde{u}$, from the partial regression (\ref{model-partial-iv}), to compute the covariance matrix of $\tilde{b}_1$ using (\ref{hc2}).
		\end{enumerate} 
		
		The four step procedure outlined above can be used for HC3, HC4 variants of HCCM, and might even be applied to different variants of cluster-robust covariance matrices for which Theorem~3 and~4 in \citet{ding-2021} do not hold. The first step would remain unchanged; the second step would only change if any quantity other than, or in addition to, $h_{ii}$ is needed; only the third step would change significantly, where the researcher would need to use the correct expression for the relevant estimated covariance matrix, in place of (\ref{hc2}).

	\end{appendices}
	
\end{document}